\documentclass[preprint2]{aastex}

\usepackage{natbib}

\slugcomment{To appear in the Astrophysical Journal}

\shorttitle{BLAST05: Targeted Sources}
\shortauthors{Truch, M.~D.~P.~et al.}

\newcommand{\h}{^{\mathrm h}}
\newcommand{\m}{^{\mathrm m}}
\newcommand{\s}{^{\mathrm s}}
\newcommand{\IRAS}{{\it IRAS}}
\newcommand{\ISO}{{\it ISO}}
\newcommand{\Spitzer}{{\it Spitzer}}
\newcommand{\Herschel}{{\it Herschel}}
\newcommand{\E}[1]{\ensuremath{\times 10^{#1}}}

\begin{document}

\title{The Balloon-borne Large Aperture Submillimeter Telescope (BLAST) 2005:
       Calibration and Targeted Sources}

\author{
  M.~D.~P.~Truch,\altaffilmark{1}
  P.~A.~R.~Ade,\altaffilmark{2}
  J.~J.~Bock,\altaffilmark{3,4}
  E.~L.~Chapin,\altaffilmark{5}
  M.~J.~Devlin,\altaffilmark{6}
  S.~Dicker,\altaffilmark{6}
  M.~Griffin,\altaffilmark{2}
  J.~O.~Gundersen,\altaffilmark{7}
  M.~Halpern,\altaffilmark{5}
  P.~C.~Hargrave,\altaffilmark{2}
  D.~H.~Hughes,\altaffilmark{8}
  J.~Klein,\altaffilmark{6}
  G.~Marsden,\altaffilmark{5}
  P.~G.~Martin,\altaffilmark{9,10}
  P.~Mauskopf,\altaffilmark{2}
  C.~B.~Netterfield,\altaffilmark{10,11}
  L.~Olmi,\altaffilmark{12,13}
  E.~Pascale,\altaffilmark{11}
  G.~Patanchon,\altaffilmark{14}
  M.~Rex,\altaffilmark{6}
  D.~Scott,\altaffilmark{5}
  C.~Semisch,\altaffilmark{6}
  C.~Tucker,\altaffilmark{2}
  G.~S.~Tucker,\altaffilmark{1}
  M.~P.~Viero,\altaffilmark{10}
  D.~V.~Wiebe\altaffilmark{11}
}

\altaffiltext{1}{Department of Physics, Brown University, 182 Hope Street, Providence, RI 02912;
                {\url{matthew@truch.net}}}
\altaffiltext{2}{Department of Physics \& Astronomy, Cardiff University, 5 The Parade, Cardiff, CF24~3AA, UK}
\altaffiltext{3}{Jet Propulsion Laboratory, Pasadena, CA 91109-8099}
\altaffiltext{4}{Observational Cosmology, MS 59-33, California Institute of Technology, Pasadena, CA 91125}
\altaffiltext{5}{Department of Physics \& Astronomy, University of British
                 Columbia, 6224 Agricultural Road, Vancouver, BC V6T~1Z1,
                 Canada}
\altaffiltext{6}{Department of Physics and Astronomy, University of Pennsylvania, 209 South 33rd Street,
                 Philadelphia, PA 19104}
\altaffiltext{7}{Department of Physics, University of Miami, 1320 Campo Sano Drive, Coral Gables, FL 33146}
\altaffiltext{8}{Instituto Nacional de Astrof\'{\i}sica \'Optica y Electr\'onica (INAOE),
                  Aptdo. Postal 51 y 72000 Puebla, Mexico}
\altaffiltext{9}{Canadian Institute for Theoretical Astrophysics, University of Toronto,
                 60 St. George Street, Toronto, ON M5S~3H8, Canada}
\altaffiltext{10}{Department of Astronomy \& Astrophysics, University of Toronto, 50 St. George Street,
                  Toronto, ON  M5S~3H4, Canada}
\altaffiltext{11}{Department of Physics, University of Toronto, 60 St. George Street, Toronto, ON M5S~1A7, Canada}
\altaffiltext{12}{Istituto di Radioastronomia, Largo E. Fermi 5, I-50125, Firenze, Italy}
\altaffiltext{13}{University of Puerto Rico, Rio Piedras Campus, Physics Dept., Box 23343, UPR station,
                  San Juan, Puerto Rico}
\altaffiltext{14}{Laboratoire APC, 10, rue Alice Domon et L{\'e}onie Duquet 75205 Paris, France}

\begin{abstract}
The Balloon-borne Large Aperture Submillimeter Telescope (BLAST) operated
successfully during a 100-hour flight from northern Sweden in June 2005
(BLAST05)\@.  As part of the calibration and pointing procedures, several
compact sources were mapped, including solar system, Galactic, and
extragalactic targets, specifically Pallas, CRL~2688, LDN~1014,
IRAS~20126+4104, IRAS~21078+5211, IRAS~21307+5049, IRAS~22134+5834,
IRAS~23011+6126, K3-50, W~75N, and Mrk~231.  One additional source, Arp~220,
was observed and used as our primary calibrator. Details of the overall BLAST05
calibration procedure are discussed here.  The BLAST observations of each
compact source are described, flux densities and spectral energy distributions
are reported, and these are compared with previous measurements at other
wavelengths.  The 250, 350, and 500\,\micron\ BLAST data can provide useful
constraints to the amplitude and slope of the submillimeter continuum, which in
turn may be useful for the improved calibration of other submillimeter
instruments.
\end{abstract}


\section{Introduction}     \label{sec:intro}
The June 2005 flight of the Balloon-borne Large Aperture Submillimeter
Telescope (BLAST) incorporated a 2-m spherical primary mirror and large-format
bolometer arrays operating at 250, 350, and 500\,\micron.  A complete
description of the BLAST instrument is given in \citet{pascale2008}.  The BLAST
wavelengths sample the peak of the spectral energy distribution (SED) for cool
dust (${\sim}\,10$--40\,K) and are in a regime which is difficult or impossible
to access from even the best ground-based sites.  As a result, BLAST has the
ability to conduct unique Galactic and extragalactic submillimeter surveys with
arcminute resolution and high sensitivity.  BLAST's primary scientific
motivations are to study the angular and redshift distribution and evolution of
high-redshift star-forming galaxies and to identify proto-stellar cores and the
earliest stages of star formation within Galactic molecular clouds.

BLAST conducted a 100-hour flight, launching from northern Sweden on 2005 June
12, and landing in northern Canada on 2005 June 16 (BLAST05)\@.  One
relatively shallow
extragalactic field and several large (${\sim}\,8\,{\rm deg}^2$) Galactic fields
\citep[see][]{chapin2008, hargrave2008} were mapped, from which a large number
of Galactic pre-stellar sources were extracted at very high signal-to-noise
ratio.  The results obtained from the BLAST flight from Antarctica in December
2006 (BLAST06) will be discussed in future articles.

BLAST05 continuum observations were made with three focal plane arrays
consisting of 139, 88, and 43 detectors at 250, 350, and 500\,\micron,
respectively.  The optics and detector layout in the cryostat make 
simultaneous measurements possible by three arrays, having a common field of view of
about 14\arcmin\ $\times$ 7\arcmin.  The detectors are silicon-nitride
micromesh (``spider-web'') bolometric detectors coupled with $2f\lambda$ 
\citep[maximum optical efficiency,][]{griffin2002}
spaced feedhorn arrays \citep{turner2001}.  The arrays are prototypes of those
developed for the SPIRE instrument on \Herschel\ \citep{griffin2004}.

In this paper we report the observations of compact sources that were targeted
by BLAST largely for the purposes of flux and pointing calibration.  These
sources include Pallas, CRL~2688, K3-50, W~75N, Mrk~231, and Arp~220, which have
all been well studied in the submillimeter and are commonly used as primary or
secondary standards.  We also conducted several observations of known bright
proto-stars that were not otherwise included in our wider Galactic plane survey
fields, specifically IRAS~20126+4104, IRAS~21078+5211, IRAS~21307+5049,
IRAS~22134+5834, and IRAS~23011+6126.  Finally we observed the ``starless''
dense core LDN~1014 that was recently studied with the {\it Spitzer Space
Telescope}\ \citep{young2004}.

Together these observations represent a broad sampling of bright submillimeter
sources that were available during the flight, given the strict visibility
constraints of the telescope \citep{pascale2008}.  The planets Mars and Uranus,
often used as submillimeter calibrators, were not visible during this flight.
Although observations of Saturn were attempted, there were concerns about gain
stability because it was relatively close to the direction of the Sun, and it may
also have suffered from saturation effects in the detectors, and hence we
abandoned attempts to use these particular observations.  After carefully
assessing the systematic effects in the relevant parts of the SEDs of the other
sources, we chose to single out Arp~220 as our primary flux calibrator.
Using Arp~220, we determine that the BLAST05 calibration uncertainties are 12\%,
10\%, and 8\% in the 250, 350, and 500\,\micron\ bands, respectively.

This paper is divided into three main sections.  Section~\ref{sec:red}
outlines the basic reduction steps and characterization of BLAST05
data.  Section~\ref{sec:perf} discusses the performance of the warm
optics in BLAST05 and the impact of the degraded performance on the
resolution and sensitivity of the experiment.
Section~\ref{sec:fluxcal} describes in detail the absolute calibration derived from  
the primary flux-calibrator Arp~220, while Section~\ref{sec:obs}
summarises the BLAST flux-densities and the simple fits of a
modified black-body emission model to the SEDs of each of
the targeted sources from the BLAST05 flight.

\section{Data Reduction} \label{sec:red}
We now describe the main steps in the data reduction and calibration
process.  Raw data from BLAST consist of a set of bolometer timestreams in
voltage units, sampled at $100\,$Hz.
The raw bolometer data are first cleaned for post-flight analysis.  The
data are de-spiked and then deconvolved to remove the electronics filters from
the timestreams; see \citet{pantachon2008} for specifics of the cleaning and
deconvolution performed on the BLAST data.  The cleaned data are combined with
a post-flight pointing solution \citep{pascale2008} to make maps at each
frequency, taking advantage of the multiple detectors, as well as significant
scan cross-linking, to minimize striping due to instrumental drifts.

The bolometers in each array are corrected for relative gains, or flat-fielded, so that meaningful
multi-bolometer maps can be generated.  The flat-field corrections are
determined using individual maps made for each bolometer from a single
point-source calibrator, in this case, CRL~2688 (the brightest point-like source observed).  
These scans were designed
such that a fully-sampled map can be generated from each bolometer
individually.  For each bolometer, the total flux from the point source is
integrated and a flat-fielding coefficient for each bolometer is then calculated
as the ratio of the flux from that bolometer and the flux from an arbitrarily
chosen reference bolometer.  All subsequent maps are generated from timestreams
that apply the flat-fielding coefficient appropriately to each bolometer.

To calculate the flux density from a point source, we adopted a
matched-filtering technique similar to that used to extract point sources from
several recent extra-galactic submillimeter surveys
\citep[e.g.,][]{coppin2006,scott2006}. A point spread function (PSF) is
generated by stacking and averaging several point sources from various maps in
telescope coordinates (azimuth and elevation relative to the telescope).  
We area-normalize the PSF, $P,$ such that
\begin{equation}
\sum_{x,y} P(x,y)\delta x \delta y = 1,
\end{equation}
where the double sum is over all pixels in the PSF map, and $\delta x$ and
$\delta y$ are the angular dimensions of a pixel such that the units of $P$ are
sr$^{-1}$.  In all BLAST maps, square pixels are used so that
$\delta x = \delta y$.  BLAST maps are calibrated in surface brightness units
[Jy sr$^{-1}$].  A map of a point source, $M(x,y)$, can be modeled by the
normalized PSF centered over the source, scaled by its flux density, $S$, or
$S P(x,y) \simeq M(x,y)$.  If each pixel in $M(x,y)$ has an uncertainty
$\sigma(x,y)$, calculated in the map-making process \citep{pantachon2008}, 
then we can write $\chi^2$ for the model, assuming independant noise:
\begin{equation}
\chi^2 = \sum_{x,y}\frac{\left(S P(x,y) - M(x,y)\right)^2}{\sigma^2(x,y)}.
\end{equation}
Minimizing this $\chi^2$ results in the maximum likelihood flux density,
\begin{equation} \label{eq:flux}
S = \frac{\sum_{x,y}(M(x,y)P(x,y)/\sigma^2(x,y))}
         {\sum_{x,y}(P(x,y)/\sigma(x,y))^2}.
\end{equation}

This technique uses our knowledge of the beam as a model for the
shape of un-resolved point-sources in BLAST maps. Fitting the amplitude
of this template is optimal in a S/N sense, and particularly important
for measurements of faint sources. This method produces smaller
measurement uncertainties than simple aperture photometry, as pixels
near the peak signal are weighted more heavily than pixels in the
wings of the brightness distribution (aperture photometry weights all
pixels equally).

This technique gives a low statistical uncertainty both because we have
intrinsically high signal-to-noise data and also because knowledge of the beam
shape is used to weight the uncertainties; however, errors in the PSF will bias
the result.  To check for a bias we used simple aperture photometry as a second
measure of flux densities for CRL~2688.  The difference between the matched
filter and aperture photometry flux densities of CRL~2688 is less than 3\%.
Since the matched filter provides significantly smaller statistical errors, it
is used to extract flux densities from all point-like sources in this paper.

The PSF for each individual bolometer varies slightly across the
array.  The stacked PSF takes this effect into account for the bulk of
the map over which each bolometer contributes approximately equally to
the combined signal.  However, the extreme edges of the map are only
sampled by a fraction of the bolometers and therefore exhibit
different effective PSFs.  We extract flux densities from maps made
with single bolometers using the stacked PSF to assess the error
introduced by adopting this incorrect template, and find at most a
10\% bias.  The bias is maximized at the very edges of the map, but
affects flux densities measured to within 14\arcmin\ (width of the
array) of the edge of the map along the scan direction (approximately
aligned with R.A.), and 7\arcmin\ (height of the array) in the
transverse direction (approximately aligned with Dec.).  None of the
measurements discussed in this paper are affected by this bias.

The flat-fielding process was repeated for CRL~2688 observations made during the
middle and end of the flight to check for stability.  Although no bias or trend
is seen, the coefficients for each individual bolometer vary by 5\% rms.
This effect averages down when multiple
bolometers are used, and is negligible in full bolometer-array maps as used in
this paper.  The 5\% scatter in flatfield coefficients appears to be dominated
by measurement uncertainties; therefore this value can be considered a very
robust upper-limit on the systematic drift in the relative calibration, and is 
insignificant compared to the total calibration error budget given in
Table~\ref{calib}.

\subsection{Responsivity Variations}
Variations in bolometer loading (due to changing sky emission, for example) and
bolometer base-plate temperature cause changes in bolometer responsivities.  To
characterize and correct for these changes, a calibration lamp (``cal-lamp'') is
located in the optics box at the center of the reflective Lyot stop \citep[see
\S 3 of][]{pascale2008}, allowing for the measurement of these changes.  The
cal-lamp is pulsed once every 10--15 minutes during flight.  The cal-lamp is of
the same design as the one used for SPIRE \citep{hargrave2006}.  It is designed
to provide a stable signal over the flight, so any change in cal-lamp amplitude
as measured by a bolometer is directly proportional to a change in that
bolometer's responsivity.  The signal baseline is removed by fitting a line to
a 550\,ms segment of data before and after the cal-lamp pulse.  A template
cal-lamp profile from a raw bolometer timestream is fit to every cal-lamp pulse
in every bolometer.  The template is chosen from a typical bolometer and
it has been verified that the template fit amplitude is within 2\% of a
square-wave fit.  The amplitude of the fit is interpolated
over time and inverted to generate a multiplicative scaling cal-lamp timestream
for each bolometer.  Fluctuations in the responsivity are less than 8\% rms,
dominated by diurnal variations due to differences in atmospheric loading and
thermal emission from the telescope's changing temperatures.  On time scales of
a typical source map the fluctuations are less than 2\% rms.  A constant
responsivity timestream for each bolometer is generated by applying the
multiplicative scaling cal-lamp timestream.  The large-scale fluctuations are
suppressed and the final responsivity variations are less than 2\% rms.

\section{BLAST05 Warm Optics Performance} \label{sec:perf} 
\begin{figure*}[t]
\includegraphics[angle=0, width=6.5in]{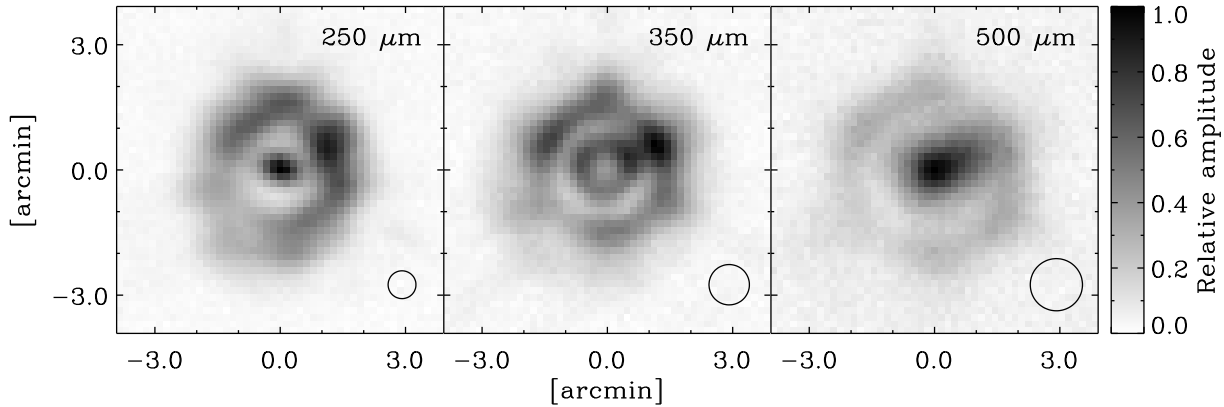}
\caption{ Point Spread Functions (PSFs), provided by observations of
CRL~2688 (\S~\ref{crl}), for each of the three wavebands from BLAST05,
generated by stacking several point source maps in telescope
coordinates.  The small circles represent the expected diffraction
limited FWHM for each of the wavebands (Pascale et al. 2008).  The
PSFs are significantly more extended than expected, reducing
sensitivity to point sources and the ability to distinguish crowded
sources amidst fluctuating cirrus.
\label{psfs}
}
\end{figure*}

During the
BLAST05 flight, the warm optics (primary and secondary mirrors) did not perform
within the specifications.  The beams as designed were expected to be close to
diffraction limited and approximately Gaussian,
$\exp(-\theta^2/2\sigma_{\rm B}^2)$,
with $\rm{FWHM} = 2 (2 \ln 2)^{1/2} \sigma_B =$ 32\arcsec,
48\arcsec, and 64\arcsec, at 250, 350, and 500\,\micron, respectively.  In this
case the full width at half power of the beam, FWHP, is equal to the FWHM, and
the beam solid angle can be characterized as
$\Omega = \pi \alpha_{\rm D}^2/(4 \ln 2)$ with $\alpha_{\rm D} = \rm{FWHP}$.
The point-spread functions in all three BLAST bands were measured from
multiple observations of the proto-planetary nebula CRL~2688.
The in-flight beam-shapes are shown in Figure~\ref{psfs}. Noting that the
central structure in each is on a scale comparable to the diffraction limit,
the beams are clearly far from ideal, distributing considerable power into an
outer ring (hexagon) of diameter ${\sim}\,200$\arcsec.

With the pre-flight predicted Noise Equivalent Flux Density (NEFD) of 200
mJy\,${\mathrm s}^{1/2}$ we would expect 1-$\sigma$ surface brightness
fluctuations at the nominal resolutions of
${\sim}\,\rm{NEFD}/\Omega $
or 11, 4.7, and 2.6 MJy sr$^{-1}\,{\mathrm s}^{1/2}$ at 250, 350, and
500\,\micron, respectively.
The measured sensitivities were in fact 8.8, 4.8, and
$2.7\,{\rm MJy\,sr^{-1}}\,{\mathrm s}^{1/2}$, respectively,
which shows that BLAST did not suffer a reduced
sensitivity to extended diffuse sources, i.e. detectors worked as planned and optical
efficiency was as planned.

There is, however, a marked loss in sensitivity for point sources, because of
the larger-than-expected beam solid angle.  The FWHP of the degraded beam is
${\sim}\,200$\arcsec\ (somewhat smaller at $500\,\mu$m), given the power in the
ring.  The reduction in point-source sensitivity is
${\lesssim}\,(200/32)^2\,{\simeq}\,40$ at
250\,\micron.  This is a pessimistic estimate because there is
considerable higher-resolution structure in the beam, which one takes advantage of in the
PSF fitting described above.  Alternatively, one can examine 
the histogram of pixel brightnesses contributing to the PSF, and
working from the brightest, find the beam area that accounts for half the power.
We find $\alpha_{\rm D}
= $ 186\arcsec, 189\arcsec, and 189\arcsec\ for 250, 350, and 500\,\micron,
respectively, and so the sensitivity is reduced by factors of
approximately 34, 16, and 9.
Despite this reduction, the targeted
sources described in this paper were still detected with high signal to noise.
Further data from BLAST05 are discussed in \citet{chapin2008},
\citet{hargrave2008}, and other forthcoming papers.

Although it is not critical to the analysis here, we have investigated the
origin of the beam degradation.  One possibility is misalignment,
shifting, and tilting of the secondary and primary with respect to the nominal
optical axis. The launch was sufficiently violent to cause 
the inner-frame locking mechanism to fail; the inner-frame (which houses
the primary, secondary, and camera) violently hit both hardware
elevation stops, which could have damaged the telescope, including the
secondary support mechanism and the primary itself.
From visual inspection of the images (and power spectra) there is clearly
information in the beams near the diffraction limit, such as the central dip at
350\,\micron, which is not present at the other two wavelengths, and the width
of the outer-ring.  While these structures are suggestive a problem in the telescope, 
numerous experiments with the
optical design software ZEMAX\footnote{\url{http://www.zemax.com/}}, in which
all the optical elements were shifted and/or tilted from their nominal
position, we were unable to reproduce the distinctive features in the beams
at all the wavelengths simultaneously.
There appears to have been an optical problem beyond  out-of-focus or 
out-of-alignment optics.
The hexagonal shape of the beam does suggest damage to the
carbon-fiber primary itself, which was constructed from six panels
(segments).  In addition to effects of the violent launch, another contributing 
factor could have been the light rain the payload endured before
launch.  Carbon fiber becomes particularly weak in a high humidity
environment, possibly leading to delamination of the six panels either
before or during the launch.  Furthermore, there was the usual
freeze-thaw cycle on ascent, where the ambient temperature dropped as
low as $-65$\,\degr{}C.
Further analysis of the PSF (including Fourier Transform) was not performed as
phase-information is required for proper analysis; as the mirror was fully
destroyed upon landing there is little use in debugging the system further.

\section{Astronomical Flux-calibration} \label{sec:fluxcal}

The primary scientific goals of the BLAST experiment demanded a flux
calibration-accuracy of better than 10\% in all three BLAST
pass-bands. Achieving this was complicated by the variable
sky-visibility of BLAST due to the unstable projected-latitude of the
telescope gondola during the flight, and the restrictions on
visibility due to the Sun and Moon avoidance criteria, the orientation
of sun-shields and other baffling, and the elevation range
(25--60\degr{}) of the gondola's inner-frame
\citep{pascale2008}. Consequently BLAST had only limited access to the
calibration sources commonly used at submillimeter and far infrared (FIR) 
wavelengths.

Since the ecliptic-plane was not visible during the BLAST05 flight, no
absolute flux-calibration of BLAST could be determined from
observations of Uranus or Mars, for which model SEDs are known to have
systematic uncertainties ${<}\,5$\% at submillimeter wavelengths
\citep{griffin1993,wright2007}.  The pre-flight strategy for achieving
a 10\% calibration accuracy, recognising the above visibility
constraints, therefore forced us to identify other Galactic and
extragalactic sources that could act as primary and secondary
calibrators, with the following requirements: (i) availability
throughout the flight; (ii) already considered, in some cases, as
secondary calibrators for ground-based sub-mm telescopes and FIR
satellites; (iii) well-constrained SEDs in the FIR to mm-wavelength
regime, enabling accurate interpolation of the band-averaged
flux-densities at BLAST wavelengths; (iv) bright ($\gg 1$~Jy at
500\,\micron) and compact sources (with respect to the BLAST beam-size,
$< 20$\arcsec) that reside in regions with minimal spatial-structure
in the Galactic foregrounds or backgrounds, allowing accurate
subtraction of any extended emission.  Given these criteria we
scheduled regular observations throughout the flight of bright
embedded protostellar-sources and compact \ion{H}{2} regions within the W58
and Cygnus-X molecular-cloud complexes (including K3-50, W~75N, and
DR21), a post-AGB star with planetary nebula (CRL~2688), the asteroid
2~Pallas, and two ultraluminous infra-red galaxies (ULIRGs), Arp~220 and Mrk~231.

In the following sub-section we justify our selection of the ULIRG
Arp~220 as our primary calibration source for the BLAST05 flight.

\begin{figure}[t]
\begin{center}
\includegraphics[width=3in]{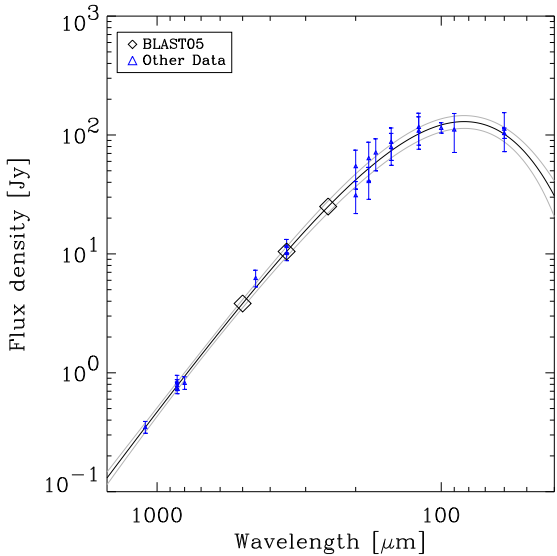}
\caption{ Spectral Energy Distribution (SED) of Arp~220, the absolute
flux calibrator for BLAST05. The best-fit model (heavy solid-line) was
constrained by the published data (blue triangles, discussed in
Section~\ref{sec:arp220}), and excluded BLAST05 measurements. The
grey-lines shows the 68\% confidence interval, estimated from 100
Monte-Carlo simulations, about the best-fit model. Black diamonds
indicate the model predictions for BLAST05 at 250, 350 and
500\,\micron. The uncertainties associated with these predictions are
11\%, 10\%, and 8\% in each band, respectively.  The best-fit
parameters of the single-temperature dust model are given in
Table~\ref{fits}.
\label{sed:arp220}
}
\end{center}
\end{figure}

\subsection{Arp~220 - a primary flux-calibrator for BLAST05} \label{sec:arp220}
Arp~220 \citep{arp1966} is a well-studied 
ULIRG at redshift $z=0.018$ with a FIR
luminosity of approximately $2 \times 10^{12} $\,L$_{\odot}$.  The
sub-millimeter emission is believed to be due to dust heated primarily
by active star-formation (at a rate of 
$\sim 200 {\mathrm M}_{\odot}/{\mathrm yr}$) and to
a lesser extent by an embedded Active Galactic Nucleus (AGN). This
emission is confined to a compact ($< 2$\arcsec\ diameter) region, as
traced through interferometric imaging, by the distribution of the
1.3\,mm dust continnum, and the molecular gas in the CO(2--1)--line
\citep{scoville1997} that fuels the starburst activity.
Arp~220 resides far from the Galactic plane ($b = +53\degr$) and
extended foreground Galactic contamination as seen in the 100 \micron\ 
map \citep{schlegel1998} does not affect the measured SED\@.
Arp~220 is therefore a point-like source as observed by BLAST, as well as 
most ground-based and space-borne FIR to millimeter telescopes, making
the construction of the FIR-to-millimeter SED insensitive to the
aperture-sizes of the photometric measurements
({Figure~\ref{sed:arp220}).

The quoted measurement errors for the ground-based 350--1100\,\micron\ 
photometry of Arp~220 are ${\leq}\,$\,10\%, while the absolute
calibration uncertainties are typically 10--20\%. In the
case of shorter-wavelength FIR measurements from ISOPHOT at
60--200\,\micron\ \citep{klaas2001} and 120--200\,\micron\
\citep{spinoglio2002}, the absolute calibration of ISOPHOT, which is
estimated to be ${\sim}\,$30\%, also dominates the total photometric
accuracy.  More precise photometry, however, at 60 and 100\,\micron\
is available from the \IRAS\ Revised Bright Galaxy Sample
\citep[RBGS,][]{sanders2003}\@, with statistical uncertainties that
have been quoted at the $\sim$\,0.1\% level.
The difference between the absolute calibration of the RBGS and the
original IRAS Bright Galaxy Samples ranges from 5--25\%, and our
analysis therefore assumes an additional 10\% absolute calibration
uncertainty on the \IRAS\ flux densities.  As we show below, combining
all the above data, and accounting for systematic uncertainties, we
have been able to constrain a simple SED model of thermal dust
emission that allows interpolated band-averaged flux-densities of
Arp~220 to be estimated across the BLAST bandpasses with an acceptable
$\sim$\,10\% uncertainty (Figure~\ref{sed:arp220}).

In order to predict the BLAST band-averaged fluxes of Arp~220, a
single-temperature modified-blackbody SED of the form $S_\nu = A
(\nu/\nu_0)^\beta B_\nu(T)$ was first fit to existing submillimeter
and FIR observations described previously.  In the above expression,
$S_\nu$ is the flux density, $A$ is the amplitude, $\beta$ is the
emissivity-index of the radiating dust-grains, $\nu_0$
is fixed at $c / 250$\,\micron, 
and $B_\nu(T)$ is the Planck function for a
blackbody with temperature $T$.

The smooth SED is fit to the data using $\chi^2$ minimization.
In addition to the measurement errors quoted in \S
\ref{sec:arp220}, we have assumed an additional {\it correlated}
error of 5\% for all of the photometry, since the bulk of these instruments
were calibrated using the same Uranus SED, which is known to have an
uncertainty of $\sim$\,5\% \citep{griffin1993}.
Furthermore, we have been careful to account for correlated errors using
the full data covariance matrix,
\begin{equation}
  \chi^2 = (\tilde{{\mathbf S}} - {\mathbf S}) C^{-1}
  (\tilde{{\mathbf S}} - {\mathbf S})^T,
\label{eq:chisquared}
\end{equation}
where ${\mathbf S}$ is the vector of measured flux densities, and
$\tilde{{\mathbf S}}$ are the model predictions.  The diagonal
elements of $C$ give the variances, $\sigma_i^2$, calculated as the
quadrature sums of the photometric and calibration uncertainties
described above.  The off-diagonal elements of $C$ are calculated as
follows.  For each pair of measurements taken with the same group and
same instrument, $i$ and $j$,
calibration uncertainties are assumed to be completely correlated such
that $C_{ij} = \sigma_i^\mathrm{c} \sigma_j^\mathrm{c}$, where
$\sigma^\mathrm{c}$ is the standard deviation of the calibration error
component for the measurement.  Monte Carlo simulations are performed
to characterize the range of models that are consistent with the data.
One thousand mock data sets are generated by adding realizations of
correlated Gaussian noise described by $C$ to the measured flux
densities ${\mathbf S}$.  Each one of these data sets is then re-fit
using Equation~\ref{eq:chisquared}.  The best-fit model ($T=41.7$\,K 
and $\beta = 1.3$, 
for the observed SED, and
its uncertainties, are shown in Figure~\ref{sed:arp220}.

The band-averaged flux densities for each BLAST wavelength are obtained for
each SED model fit as
\begin{equation}
{\tilde S_{\mathrm B}} = \int T(\nu) {\tilde S}(\nu) d\nu,
\end{equation}
where we use tildes to represent model quantities.  Here the normalized filter
transmission profiles $T(\nu)$ are measured empirically using a Fourier
transform spectrometer \citep{pascale2008}.  From these 1000 sets of
band-averaged fluxes, we obtain estimates of variances of the calibrator
brightnesses in each band, $\left<\sigma^2_i\right>$, as well as the correlated
error between the fluxes in each band, $\left<\sigma_i \sigma_j\right>$.

Maximum-likelihood estimates for the flux densities of all BLAST05
sources are obtained from equation~(\ref{eq:flux}) using un-calibrated
maps of Arp~220 in units of Volts.  Multiple observations of CRL~2688
were made throughout the flight, allowing us to estimate systematic
gain variations not traced by the internal cal-lamp pulses at the 3\%
level.  Since this gain variation is significantly lower than the
calibrator flux-uncertainty estimated above, we calculate single
calibration coefficients $c_i$ in each band for the entire flight as
the ratio of the band-averaged fluxes to the raw fluxes from our maps.
Note that this calibration depends explicitly on the assumption of a
smooth thermal SED that neglects the effects of molecular emission
lines. 

This problem has been examined in detail for SCUBA 850\,\micron\ continuum
observations of a variety of Galactic objects.  \citet{friesen2005}
conclude that the line contamination is typically $<10$\% in hot
Galactic molecular cores, and a similar conclusion was reached by
\citet{johnstone2003} in a study of protostellar sources in Orion,
although in some exceptional cases the line-contamination of the
850\,\micron\ continuum fluxes in the most energetic sources reached
levels $>50$\%. Similarly, line-contamination, due exclusively to
CO(3--2) emission, has been measured to be $<$10\% in 850\,\micron\ 
continuum fluxes, for $\sim$\,60\% of the 45 local galaxies observed by
\citet{Yao2003}.  Unfortunately there is almost no information on the
strengths of possible line-contamination in the BLAST pass-bands,
including redshifted molecular-lines at $\ll 300$\,\micron, and thus we
assume that $<10$\% line-contamination is present in the BLAST
data.

Including all of these sources of uncertainty, the final percentage
errors (8--12\%) in the absolute flux calibration, and a Pearson
correlation matrix showing the relationship between errors in the
different bands, are given in Table~\ref{calib}.  The calibration
coefficients are used to convert raw (Voltage) maps into calibrated
(Jy) maps. All astronomical analysis is performed on the calibrated
maps.

\section{BLAST05 observations of bright sources} \label{sec:obs}

In this section we summarise the calibrated BLAST observations of all
the bright targets presented in this paper, paying attention to the
level of agreement between their calibrated BLAST fluxes and the
predictions from an interpolation of their measured SEDs.  Although
these bright-sources
were selected to provide checks on the consistency of the calibration
throughout the flight, and were not considered to be part of the
scientific program, we derive some of the physical properties
(temperatures, FIR luminosities, dust masses) that can be obtained
from a simple
model of thermal dust-emission that has
been fit to the improved spectral coverage of their SEDs (\S~\ref{sec:onetemp})

In addition to conducting large-area surveys of Galactic fields
undergoing active starformation \citep{chapin2008}, 
BLAST05 observed a total of 9 bright, individual point-like sources
(Pallas, CRL~2688, IRAS~20126+4104, IRAS~21078+5211, IRAS~21307+5049,
IRAS~22134+5834, IRAS~23011+6126, Mrk~231, and Arp~220), complemented
by observations of more extended star-forming structures towards W75~N
in Cyg~X and K3-50 in W~58, as part of the overall calibration
strategy.  Although not used explicitly as a calibration source, we
also present in this paper the BLAST05 observations of the compact low-mass
protostar L1014-IRS towards the nearby Galactic dark-cloud LDN~1014.

All BLAST05 calibration targets are in the northern hemipshere
(23\degr $<$ Dec $<$ 61\degr; Table~\ref{fluxen}).  BLAST therefore
provides valuable new photometric data at intermediate wavelengths to
complement those obtained in the FIR and the (sub-)millimeter regime.

With the advantage of simultaneous observations at 250, 350, and
500\,\micron, BLAST measures the thermal emission of these Galactic and
extragalactic objects with a high flux-calibration accuracy in a
wavelength-regime that traces the spectral curvature at wavelengths
longer than the spectral peak, and shorter than the Rayleigh-Jeans
limit for cold dust ($<50$\,K). In terms of modeling the SEDs, this
short-submillimeter wavelength-regime sampled by BLAST is critical for
unlocking the degeneracies between the effect of dust-temperature,
opacity and emissivity-index on the overall FIR-to-millimeter
emission. The SPIRE instrument, using similar bolometer-arrays and
identical filters to BLAST, following the future launch of {\it Herschel}
\citep{griffin2004}, will extend this work with greater sensitivity
and spatial resolution.

It is not the purpose of this paper to provide a detailed account of physical
properties and nature of each bright-source observed in BLAST05.  Thus we only
discuss briefly in the following sub-sections the collective results btained
from these BLAST observations at 250, 350, and 500\,\micron.

\subsection{Spectral Energy Distributions}

Flux densities at 250, 350, and 500\,\micron for all the above
point-sources are extracted using the matched-filter technique
(eq.~[\ref{eq:flux}] outlined in \S~\ref{sec:red}), while for
extended or blended-sources at the resolution of BLAST05 (e.g. towards
K3-50 and W~75N), we convolve the PSF with a simple model which is then fit
to the data.  In the latter case the model consists of multiple
Gaussian sources, where we allow the amplitude, width, and position of
the Gaussians to vary as parameters of the fit.

Since the BLAST filters have large spectral widths, and colors sampled
by the filters are a strong function of temperature (especially for
$T\la25$\,K), a correction must be made to calculate monochromatic
flux densities.  Either a color correction may be applied to the
band-averaged flux densities, or effective filter wavelengths for each
object may be quoted.  We choose the former, and quote effective flux
densities at precisely 250, 350, and 500\,\micron . Once the SED
has been fit to the BLAST data by minimizing $\chi^2$, to obtain the
best-fit dust temperature and emissivity-index, this is used to
calculate the correction
\begin{equation}
S(\nu) = S_{\mathrm{B}} \frac{\tilde{S}(\nu)}{\tilde{S}_{\mathrm{B}}},
\end{equation}
where $S_{\mathrm{B}}$ is the band-averaged BLAST flux
measurement, $\tilde{S}(\nu)$ is the
SED model flux density (evaluated at 250, 350, and 500\,\micron), and
$\tilde{S}_{\mathrm{B}}$ is the SED model band-averaged flux. The
BLAST color-corrected flux-densities for all bright sources presented in
this paper are given in
Table~\ref{fluxen}.

\subsubsection{Single-temperature cold dust models}\label{sec:onetemp}
Incorporating the BLAST measurements into the existing FIR--mm
wavelength ($\sim$ 50--2000\,\micron) SEDs for each source,  we
fit only a single-temperature dust component while recognising that
any model that utilizes shorter-wavelength mid-IR data must
naturally include thermal emission from hot dust ($>$\,100\,K).  This
decision is justified in the context of deriving the band-averaged
BLAST fluxes, which require only a smooth and accurate fit to the SED
over the restricted wavelength-range of the BLAST filters
\citep{pascale2008}.  Although alternative and more complex models
(e.g.  two-temperature components) can be fit to the same FIR--mm
data, the difference in the quality (e.g. minimised $\chi^2$) of
the fit is not significant 
when compared to the single-temperature model.

As previous studies of the FIR-mm SEDS of local galaxies
and ULIRGs have shown (e.g. Eales et al. 2000, Lisenfeld, Isaak \&
Hills 2000), relying on limited photometric data, with typically 3 to
5 measurements in the FIR (from {\it IRAS} and {\it ISO}) and 
(sub-)millimeter data (at
850\,\micron\ and 1.2\,mm data), these SEDs can be well-represented by a
single dust temperature in the range of 25--50\,K.  Similar conclusions have
been reached from the studies of dust in the Galactic ISM\@.  
Even with the addition of ground-based 350 and 450\,\micron\ data
\citep[e.g, ][]{benford1999, hunter2000, dunne2001, beelen2006, coppin2008}
single temperature fits work fairly well, and there is typically no need to 
fit multiple temperature components to SEDs unless one also brings in data at 
$\lambda \ll$ 100\,\micron.  More complicated fits would only be demanded with 
higher-fidelity spectral measurements covering a much wider range of wavelengths.  

Given this, we show the mid-IR to millimeter-wavelength SEDs of all
the bright compact sources observed by BLAST05 and their best-fit
single-temperature models in 
Figs.~\ref{sed:arp220}--\ref{sed:protostellar}, and present the
best-fit model-parameters in Table~\ref{fits}.

Adopting a distance to each source (\S~\ref{sec:sources}), we determine the
FIR luminosity and dust mass from the SED fits, recognising that the
adoption of two or more temperature components would modify the
dust-mass estimate.
Determination of bolometric fluxes and dust masses follows the procedures
described in \citet{chapin2008}.
Uncertainties in quantities derived from the SED fits in
Table~\ref{fits} are obtained from Monte Carlos similar to those
described in \S~\ref{sec:arp220}, which now include BLAST data and their
correlated calibration uncertainties.  For each mock data set the
quantity in question is derived from the model fit.  These values are
then placed in histograms in order to extract means and 68\%
confidence intervals.  Further details on these Monte Carlo
simulations are provided by \citet{chapin2008}.

\begin{figure}[t]
\begin{center}
\includegraphics[width=3in]{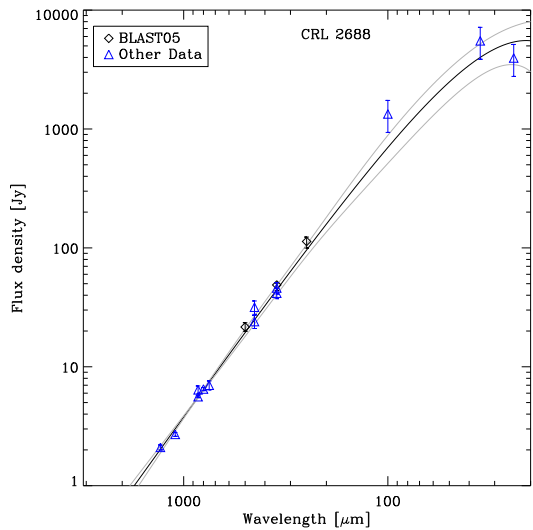}
\caption{CRL~2688 spectral energy distribution, showing BLAST data and
best-fitting models. Symbols and lines are as those described
in Fig.~\ref{sed:arp220}. \ISO~LWS data between 43--194\,\micron\ are
presented by \citet{cox1996}.  Other data are taken from
\citet{sandell1994}, \citet{jenness2002}, and \citet{omont1995}. 
\label{sed:crl2688}
}
\end{center}
\end{figure}

\subsection{CRL~2688}    \label{crl}
CRL~2688 is a post-asymptotic giant-branch star with a proto-planetary
nebula \citep[``The Egg''][]{price1983}, and it is considered an
archetypal object with which to study the evolution of AGB red giants
into bipolar planetary nebulae.  The distance to CRL~2688 is not
well-known, but is estimated  to be approximately $1.25\,$kpc
\citep{crampton1975,cohen1977}.  Despite residing in the Galactic
plane, CRL~2688 exhibits compact submillimeter emission and sufficient
contrast relative to its surroundings to provide a commonly used
secondary calibrator \citep[e.g.][]{sandell1994,jenness2002}.  The
submillimeter emission arises from a dust shell with an extent of
approximately 5\arcsec\ \citep{jenness2002}, making it point-like for
BLAST.  CRL~2688 was visited regularly throughout the flight and thus
the observations were useful for tracking the PSF shape (Fig.~2) and
relative antenna gain variations, making it our primary flat-fielding
calibrator.

Fainter surface-brightness shells around CRL~2688 have also been
observed at 120--180\,\micron with \ISO\ at a radial distance 
of $\sim$\,150\arcsec\ and 300\arcsec\ \citep{speck2000},
although this has recently been disputed with \Spitzer\ MIPS
observations \citep{do2005}.  In any case, these shells would be  large
compared to the BLAST beam, and of uniform surface-brightness, such
that any faint emission would be removed from the measured fluxes
through our baseline subtraction.  No long-term submillimeter
variability has been detected in submillimeter observations with SCUBA
\citep{jenness2002} and hence archival data can be combined to perform
our calibrations.

The mid-IR to FIR SED has been accurately measured by \ISO~LWS
between 43 and 194\,\micron\ \citep{cox1996}. At submillimeter
wavelengths, \citet{jenness2002} report the integrated fluxes
derived from SCUBA maps at 450 and 850$\mu$m, over a 40$''$ aperture
which exceed the earlier single-beam peak fluxes of Sandell
(1994). Combining these data with the new BLAST measurements
(Figure~\ref{sed:crl2688}) we find an acceptable fit to a
single-temperature modified black-body model, deriving a dust-temperature of
210\,K and emissivity-index of $0.4 \pm 0.2$. The SED of CRL~2688 in the
FIR--submillimeter regime traced by BLAST is not as accurately
constrained as that of Arp~220 as there are no FIR measurments with either
IRAS or \Spitzer\ MIPS due to saturation.  We derive uncertainties in the BLAST fluxes of
CRL~2688 that range from $\sim$\,10\% at 500\,\micron\ to $\sim$\,20\%
at 250\,\micron.

\begin{figure}[t]
\begin{center}
\includegraphics[width=3in]{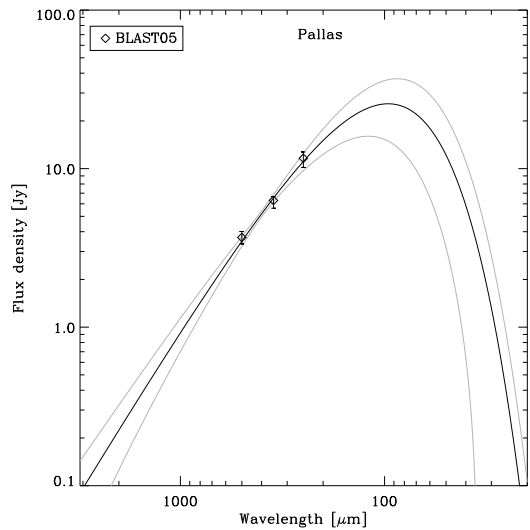}
\caption{Pallas spectral energy distribution.  Symbols and lines are as 
those described in Fig.~\ref{sed:arp220}.}
\label{sed:pallas}
\end{center}
\end{figure}

\subsection{2~Pallas}
2~Pallas is the third largest asteroid, with a diameter of 530\,km
\citep{mitchell1996}. It has a well-determined, eccentric orbit with a
semi-major axis between 2.1 and 3.4\,AU. At the time of the BLAST
observations, 2~Pallas was at a distance of 2.2\,AU\@.  With an
angular size ${<}\,1$\arcsec, Pallas is point-like to BLAST\@.  Errors
in the derived flux-densities of Pallas at BLAST wavelengths are
5--10\%, and are due mostly to uncertainties in the size, shape, and
albedo of the asteroid.  The current best estimates for the physical
properties of Pallas come from other submillimeter and IR flux density
measurements integrated into the ``thermo-physical'' model (TPM) of
\citet{muller2002}.  The TPM has an uncertainty of around 5--10\% for
integrated fluxes in the {\it ISO\/} LWS band.  The absolute
calibration of the model has a supposed accuracy of ${\sim}\,$10\% in the
longer-wavelength BLAST bands, comparing predictions with measurements
from the JCMT (850\,\micron) and CSO (350\,\micron) (T.~M{\"u}ller,
private communication).  Although this uncertainty is similar to that for the
SED of Arp~220, Arp~220 was adopted as the absolute flux
calibrator over Pallas since it is brighter, and its SED does not vary
over the duration of the flight.  Changes in apparent brightness
due to orbital effects are negligible on the timescale of the
individual BLAST observations \citep{muller2005}, and the variations
due to the 7.8~hour rotational period of 2~Pallas are less than
7\% during the entire BLAST flight (T.~M{\"u}ller, private
communication).  Although the TPM contains a full shape description,
and spin model (enabling SED predictions at any epoch), the
uncertainty in the absolute calibration of the model flux is
comparable to the predicted maximal brightness variations during the
BLAST flight.  For this reason, in Table~2 we have simply averaged
together our 4 sets of Pallas observations, and we plot only the BLAST
photometry points in Figure~\ref{sed:pallas}, with no attempt to
relate these to other data at significantly different epochs.
The TPM predicts Pallas fluxes
of 13.2, 6.8, and 3.4 Jy which are consistant with the BLAST05
measurments, differing by 14, 8.6, and 7.3\% at 250, 350, and 
500\,\micron\ respectively. The submillimeter spectral-index 
$\beta$ in the SED of
Pallas, as measured by BLAST, is $0.2 \pm 0.3$, which is the
shallowest of all the sources described in this paper.

\begin{figure}[t]
\begin{center}
\includegraphics[width=3in]{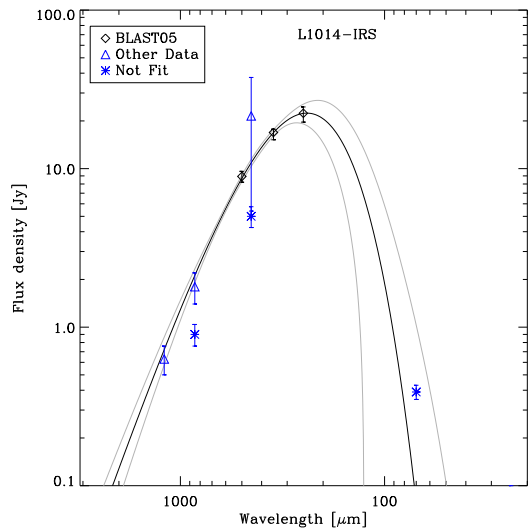}
\caption{L1014-IRS spectral energy distribution.  Symbols and lines
are as those described in Fig.~\ref{sed:arp220}. The fit is poor,
partly because a single temperature modified blackbody may not be the best
model, but also because of resolution issues with some of the
photometry.  We have not included the lower 450 and 850\,\micron\ data
points, which clearly disagree with the remaining photometry, probably
because of resolution effects.  Non-BLAST data are from
\citet{visser2002} and \citet{young2004}. 
\label{sed:l1014}
}
\end{center}
\end{figure}

\subsection{L1014-IRS}
LDN~1014 \citep{lynds1962}, which was classified as a ``starless
core'' due to the lack of a FIR \IRAS\ detection and no signature of a
continuum outflow \citep{young2004}, is a nearby dark cloud that
contains one of the lowest luminosity proto-stellar systems, L1014-IRS.
More recently, a bipolar molecular outflow has been discovered
\citep{bourke2005}. The distance to LDN~1014 is not firmly determined,
with estimates of $\sim$200\,pc \citep{huard2006}, 400--900\,pc 
\citep{morita2006} and $< 500$\,pc \citep{shirley2007}.  
The FIR--millimeter SED of L1014-IRS, derived from a physical
model of a circumstellar disk heated by the central protostellar
object, and constrained by recent \Spitzer\ observations from
3--70\,\micron\ predict a strong spectral-peak at 
$\sim$350\,\micron\ \citep{young2004}.  
Since L1014--IRS was unresolved by BLAST, we were
unable to observe the predicted density-profile. The BLAST fluxes at
350 and 500\,\micron, however, are in good agreement with the model, but
suggest that the SED peaks at shorter wavelengths, $< 250$\,\micron.

\begin{figure*}[t]
\begin{center}
\includegraphics[width=2.0in]{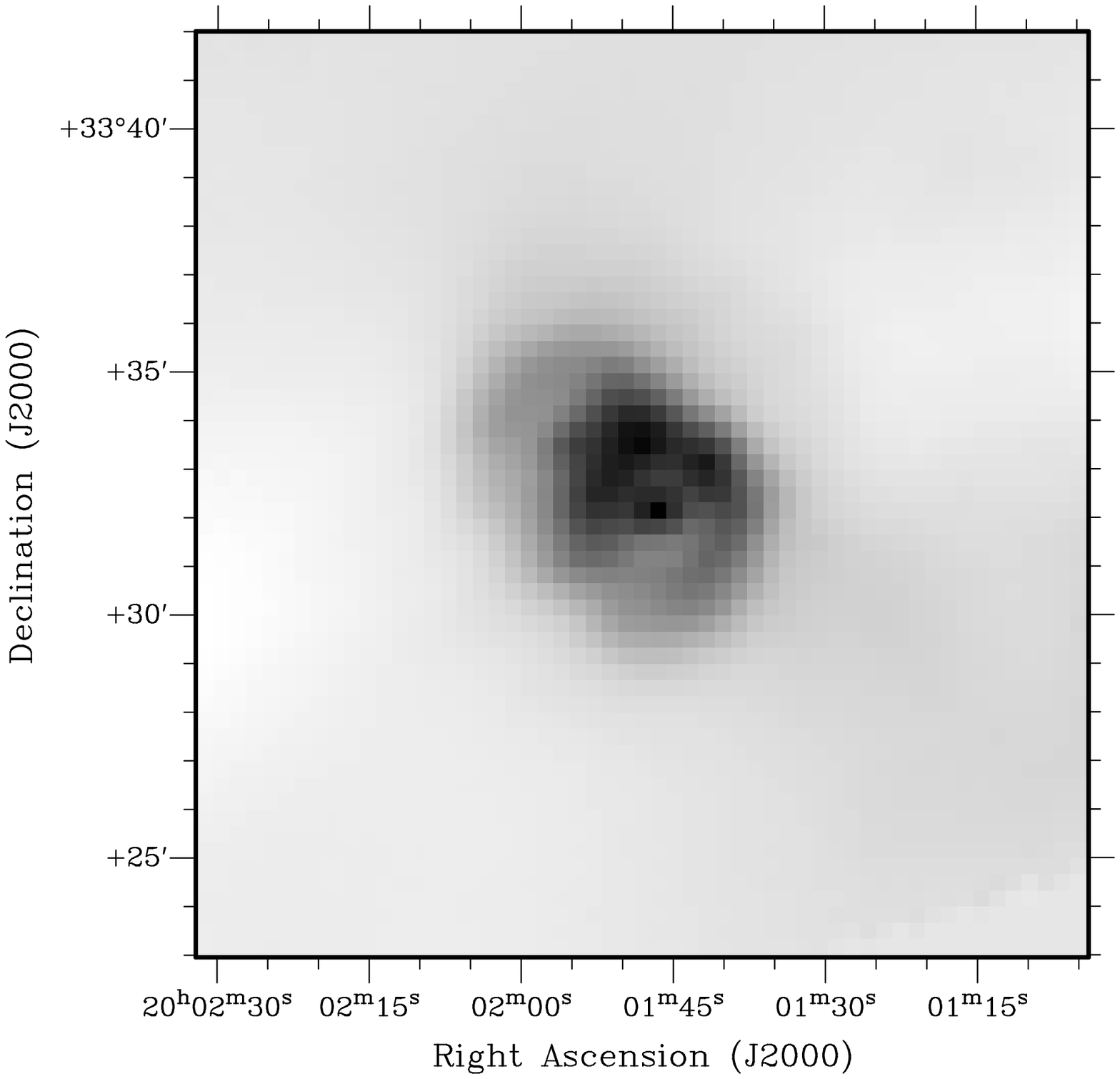}
\includegraphics[width=2.2in]{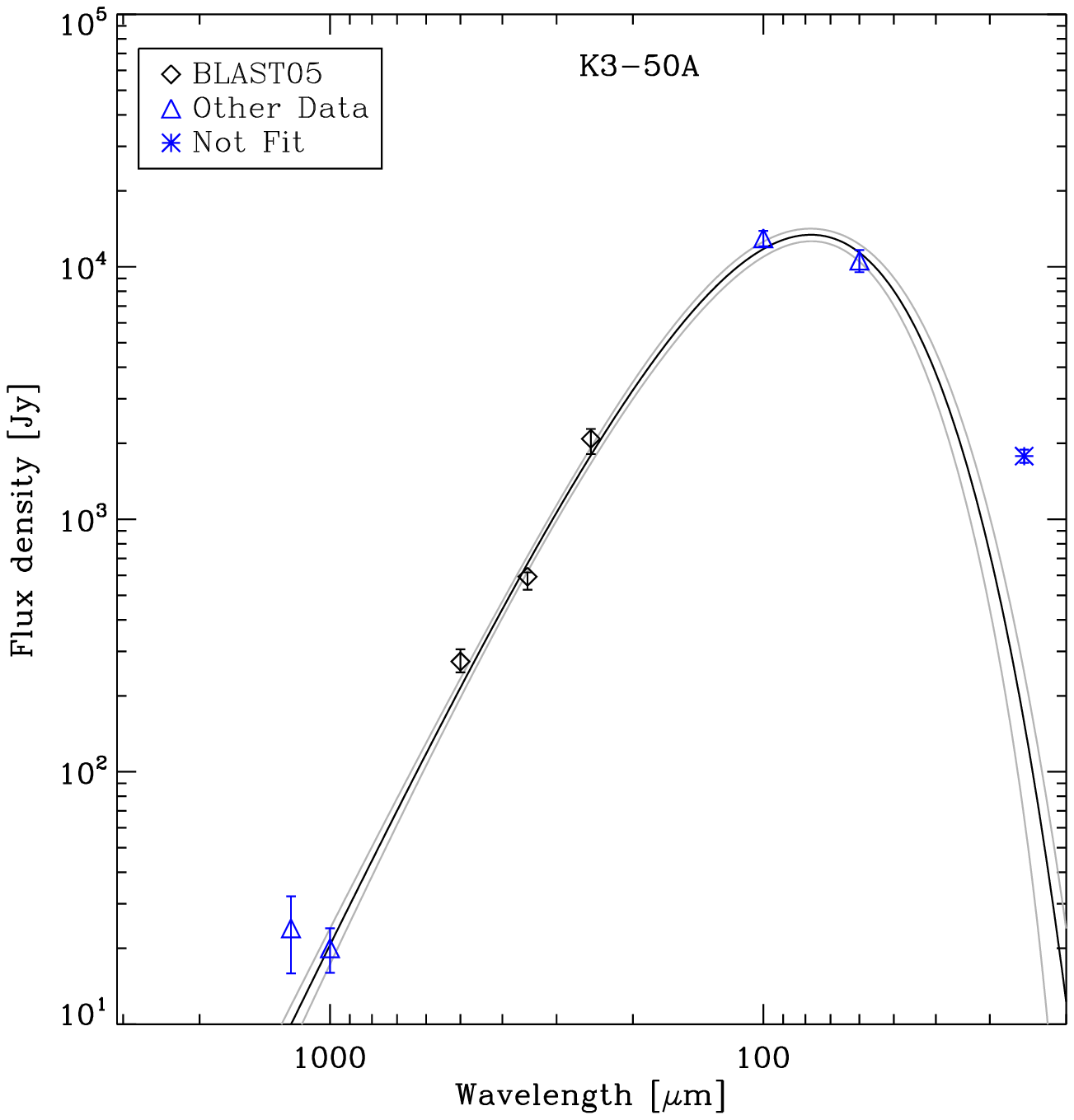}
\includegraphics[width=2.2in]{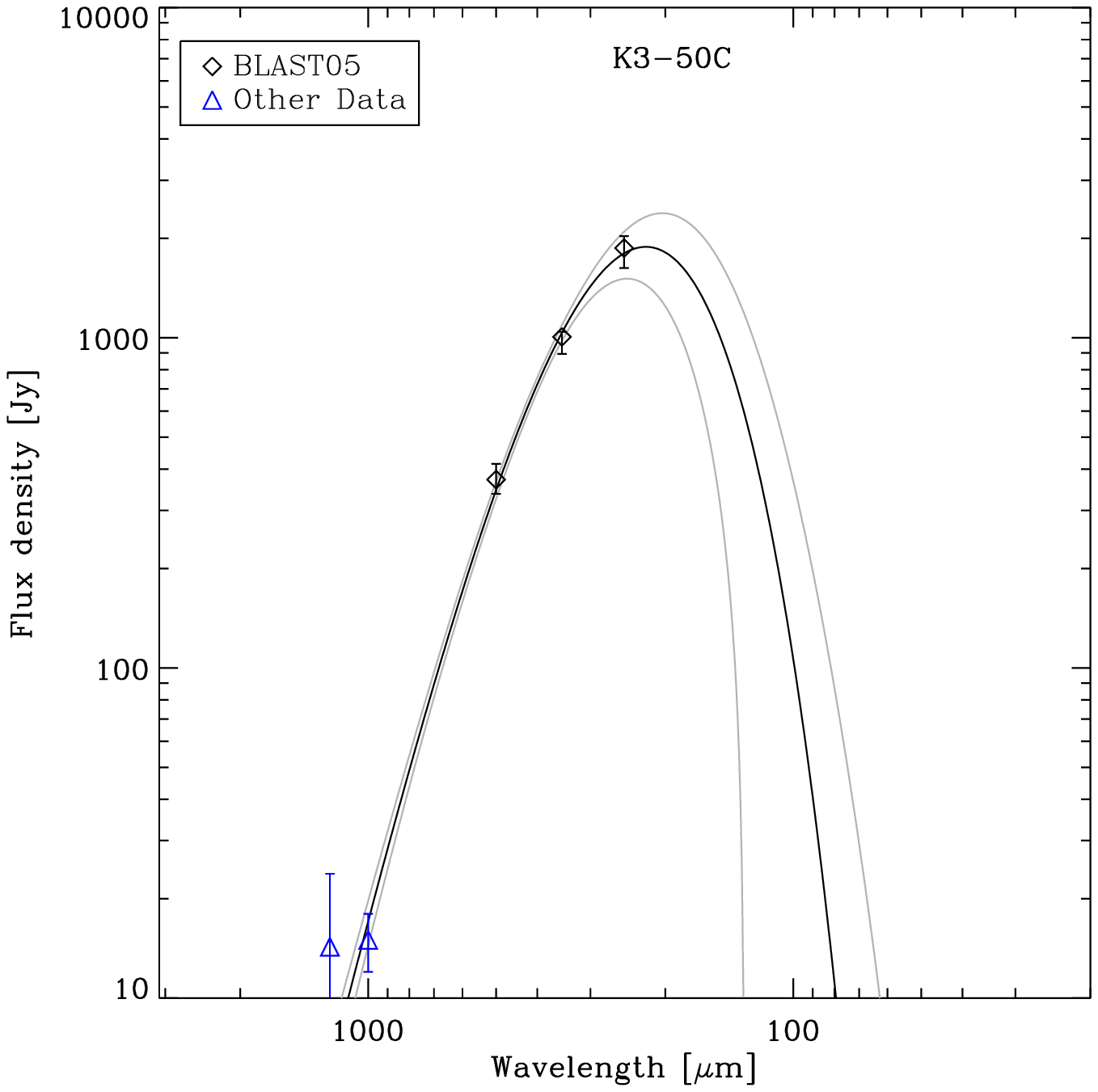}
\caption{Left: 250\,\micron\ image of the K3-50 region, with the
brighter source K3-50A to the south-west and K3-50C to the north-east.
Spectral energy distributions for K3-50A (middle panel) and K3-50C
(right panel).  Symbols and lines are as those described in
Fig.~\ref{sed:arp220}.  Other data are taken from \citet{clegg1976}
and \citet{wynn-williams1977}. The resolution of \IRAS\ makes it 
impractical to extract short wavelength flux densities for K3-50C.
\label{sed:k3-50}
}
\end{center}
\end{figure*}

\begin{figure*}
\begin{center}
\includegraphics[width=2.8in]{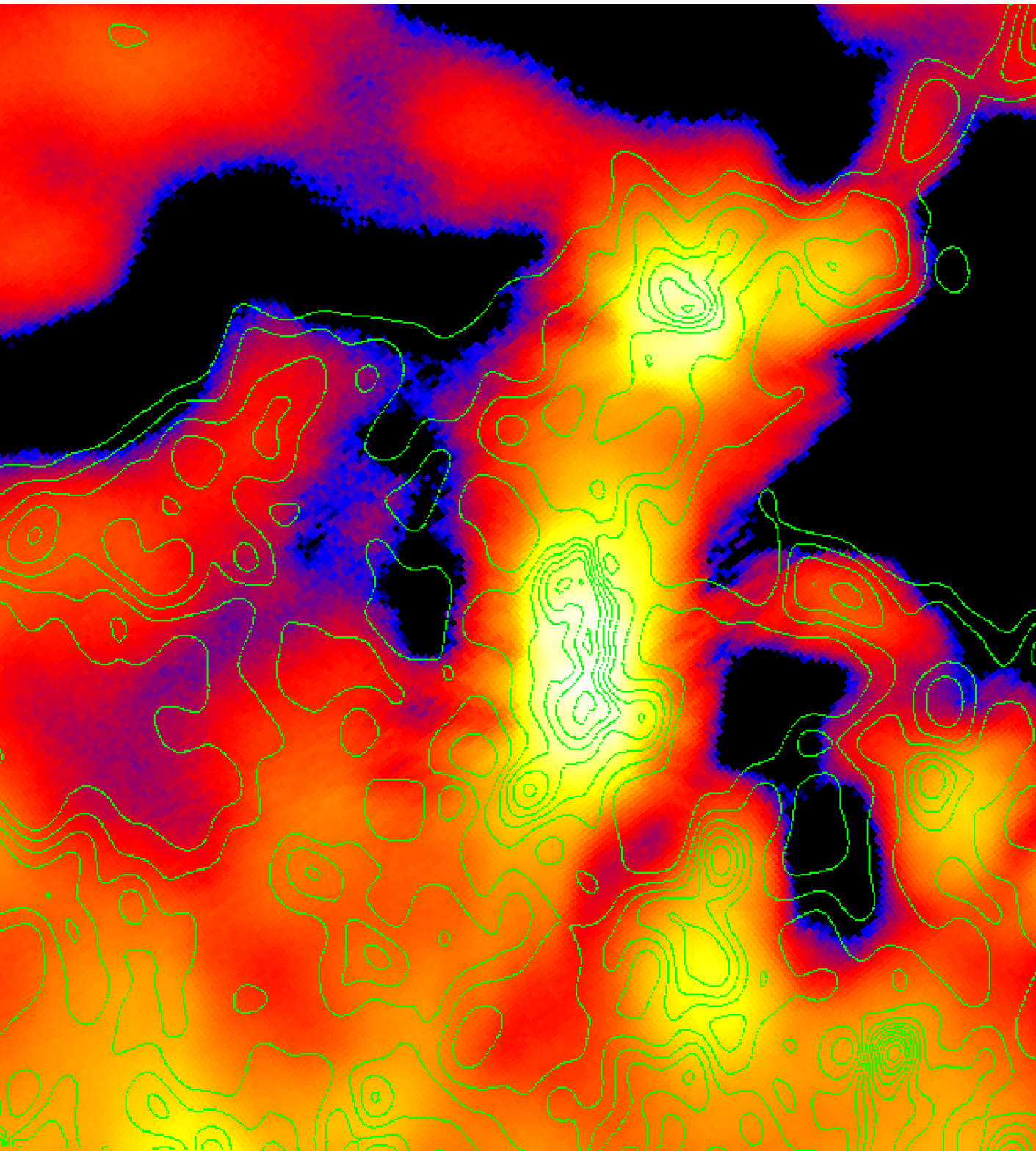}
\includegraphics[width=3in]{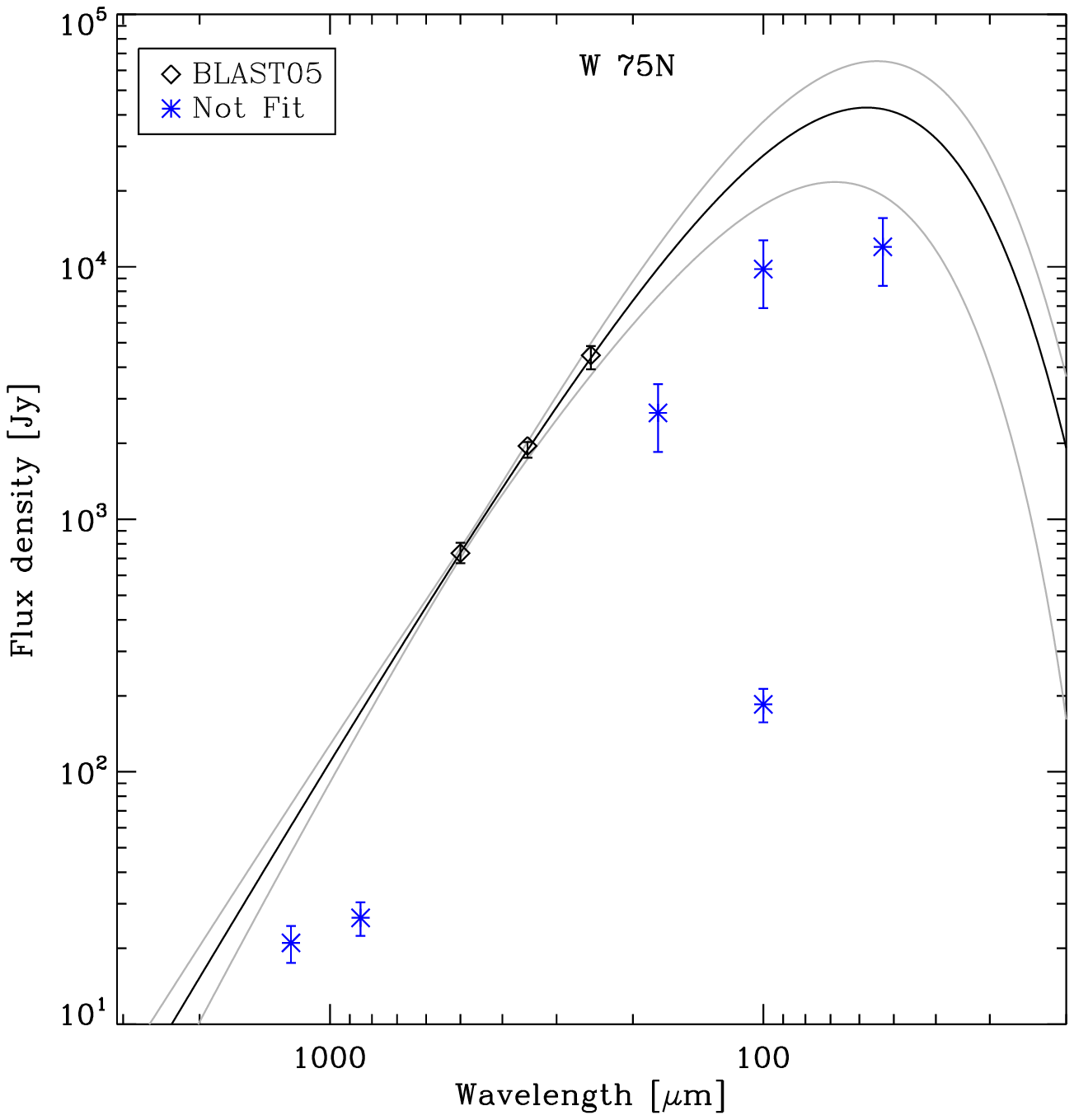}
\includegraphics[width=3in]{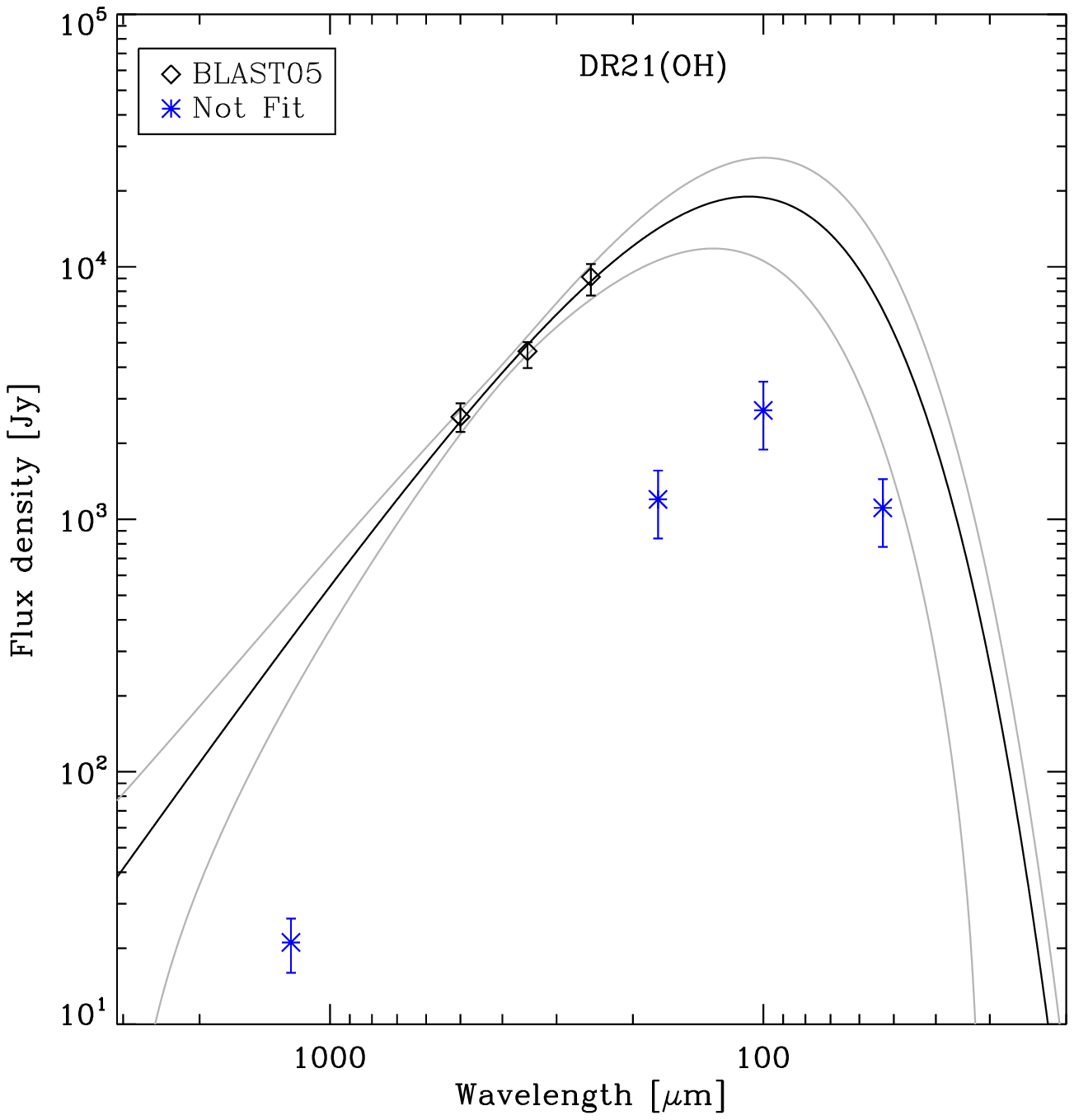}
\includegraphics[width=3in]{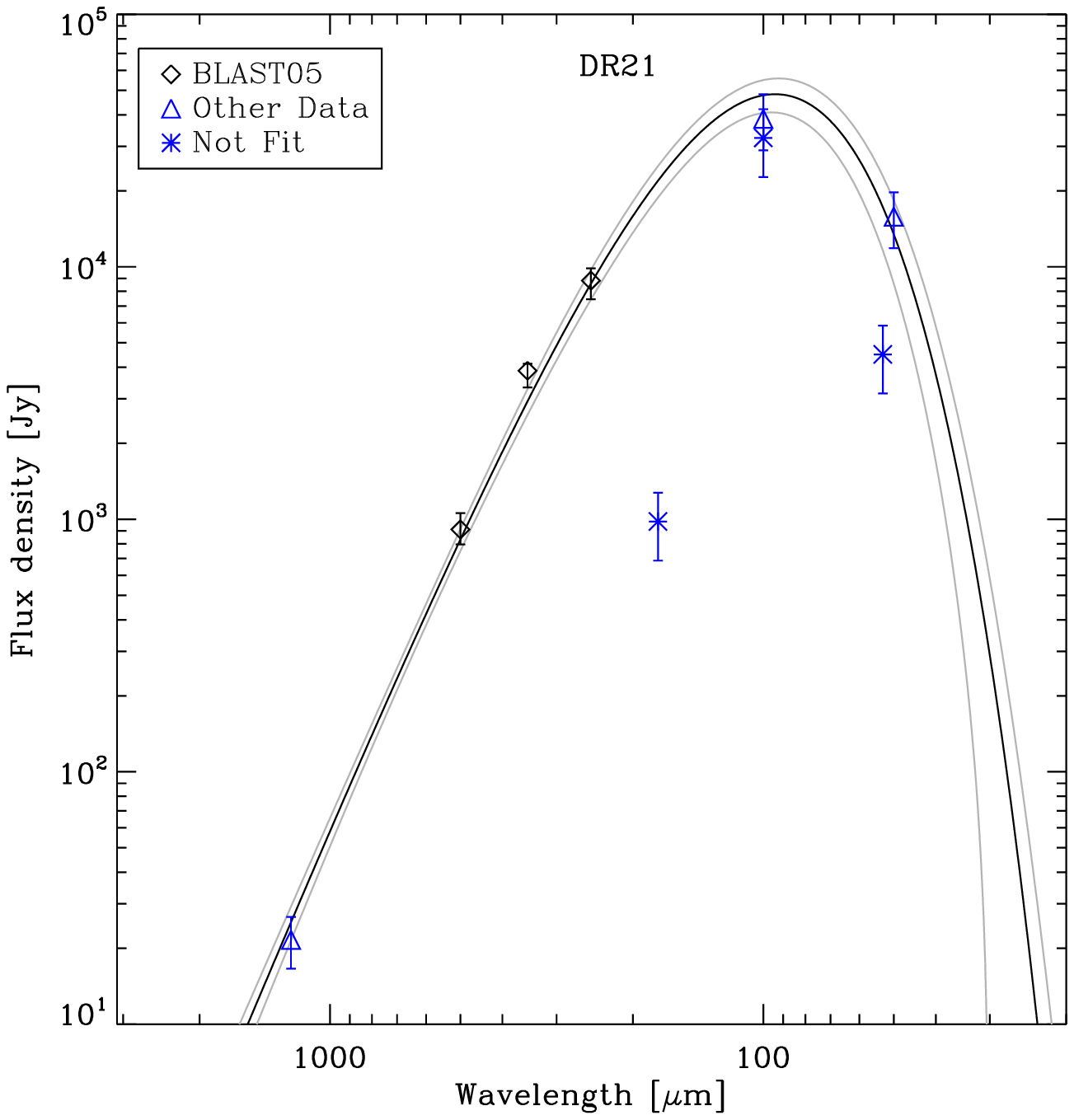}
\caption{Top-left: 250\,\micron\ BLAST image of the W~75N region with
green contours showing the distribution of IR extinction \citep{kumar2007}.  
We divide
the BLAST emission into 3 sources identified with: W~75N to the north;
DR21(OH) the northern part of the merged pair; and DR21, the
southernmost source.  The 1\arcmin\ pixel-size in this map hides much
of the structure in the BLAST PSF.  
SEDs show data and best-fitting models for W~75N (top-right),
DR21~(OH) (lower-left), and DR21 (lower-right).  Symbols and lines are
as those described in Fig.~\ref{sed:arp220}.  The data from
\citet{clegg1976}, \citet{davis2007}, \citet{helou1988}, and
\citet{harvey1977} are not adopted in the fits because of the smaller
aperture used to determine those flux density measurements.  The data
from \citet{harvey1977} suggest a temperature of 65\,K for W~75N and
40\,K for DR21~(OH), and these are used as constraints in our SED
fits.  Other data are from \citet{clegg1976} and \citet{colome1995}.
\label{sed:w75n}
}
\end{center}
\end{figure*}

\subsection{K3-50}
K3-50 is a group of compact \ion{H}{2} regions within the star-forming
complex W~58, at a distance of about 8.5\,kpc \citep{peeters2002}.
Despite the non-optimal PSF, BLAST clearly resolves at least 2 sources
separated by ${\simeq}$\,2\arcmin, which can be identified as K3-50A
and K3-50C \citep{howard1996}. Figure~\ref{sed:k3-50} shows the
250\,\micron\ image.  Fluxes for K3-50A and K3-50C are obtained by
simultaneously fitting to the BLAST data a model of two 2-dimensional
Gausssian sources convolved with the BLAST PSFs\@, whose sizes,
positions, and amplitudes are parameters in the fit.  An alternative
model of two point-sources, fixed at the positions of K3-50A and
K3-50C \citep[from][]{howard1996}, produces an indistinguishable
solution for the inferred fluxes.

\citet{thompson2006} report an integrated $450\,\mu$m SCUBA flux of
$256\pm79\,$Jy for K3-50A, which is consistent with our $500\,\mu$m
photometry, despite issues with chopping and the restricted map-size
for SCUBA making the comparison potentially complicated.  Although there
is some suggestion in the literature that these sources may be
variable\footnote{\url{http://www.jach.hawaii.edu/JCMT/continuum/calibration/sens/potentialcalibrators.html}},
we fit their individual FIR--millimeter SEDs (Figure~\ref{sed:k3-50}) with
single-temperature modified black-body emission.  K3-50C appears to have a dust temperature
of only 9\,K, substantially colder than that in K3-50A which has
dust radiating at temperatures of $\sim$40\,K
(Table~\ref{fits}), in good agreement with previous estimates
based on FIR KAO data \citep{thronson1979}.
This differece in dust-temperature may be due to the fact that K3-50C
is embedded more deeply in the molecular cloud than K3-50A, or it may
reflect the different evolutionary stages of the objects, with K3-50A
generally agreed to be younger \citep{howard1996}.

\subsection{W~75N}
The W~75N \citep{westerhout1958} and DR21 field contains a set of
young protostar/compact \ion{H}{2} regions located in the Cygnus~X
molecular cloud complex.  These sources are at a distance of about
3\,kpc \citep{campbell1982,pipenbrink1988}, although other estimates
suggest a closer distance of around $1.7\,$kpc \citep{jakob2007}.  The
region has been extensively studied over a wide wavelength range, with
many detailed spectroscopic (including OH and other maser lines) as
well as 
continuum observations
\citep[see e.g.][]{davis2007}.  Despite its extended emission,
W~75N has been suggested as a potential calibrator for submillimeter 
observations \citep{sandell2003}.  The W~75N region contains at
least 3 point-like objects which are resolved by BLAST: W~75N; DR21
\citep{downes1966}, and DR21(OH)\@.  Flux densities for W~75N are
obtained by the matched filter technique, as described in
\S~\ref{sec:red}.  Since DR21(OH) and DR21 overlap, their flux
densities are derived from a model of point sources convolved with the
BLAST PSF (as described in \S5.1), where the positions and fluxes are
the free parameters. These BLAST data suggest that DR21 has a
higher value of $\beta$ and a lower dust-temperature than DR21~(OH),
in agreement with what has been found in other studies
\citep[e.g.][]{jakob2007}. 

In addition to the detection of the protostellar cores within the \ion{H}{2} 
regions, and as can be seen in Figure~\ref{sed:w75n}, the BLAST data
also show significant extended surface-brightness structure 
($\sim 1000$\,MJy\,sr$^{-1}$ at 250\,\micron).  This is highly-correlated with the
near-IR-extinction maps towards the same region \citep{kumar2007}, indicating that it is real,
rather than being map-making artefacts.  . 

\begin{figure}[t]
\begin{center}
\includegraphics[width=3in]{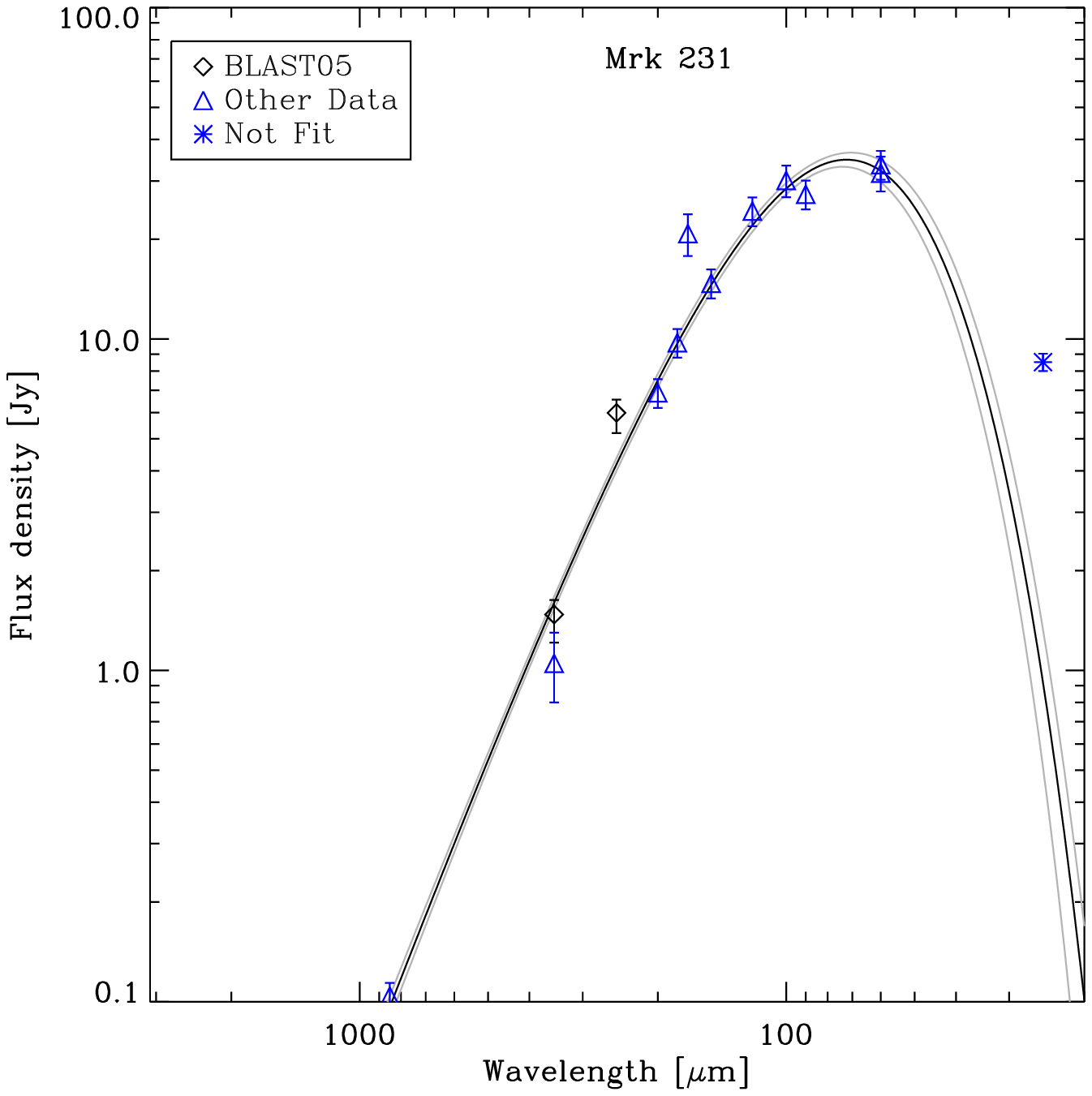}
\caption{ Mrk~231 spectral energy distribution showing data and
best-fitting model.  Symbols and lines are as those described in
Fig.~\ref{sed:arp220}.  Mrk~231 was not detected by BLAST at
500\,\micron.
Other data are taken
from \citet{benford1999}, \citet{klaas2001}, \citet{lisenfeld2000}, \citet{stickel2004},
and \citet{helou1988}.
\label{sed:mrk231}
}
\end{center}
\end{figure}

\subsection{Mrk~231}
Mrk~231 \citep{markarian1969}, also known as IRAS~12540+5708, is the
most luminous infrared galaxy within the local volume out to a
distance of 175\,Mpc.  Mrk~231 hosts a Seyfert~1 nucleus, although the
FIR emission is known to be dominated by star-formation within a
region which is of order 1\arcmin\ \citep[see][and references
therein]{lipari2005}.  It is therefore point-like for BLAST, given
the measured PSF (Fig.~1). Mrk~231 has a well measured SED and has
been used as a template for photometric redshift estimates of extragalactic
sources \citep[e.g.][]{klaas2001, aretxaga2005}.  There are useful
far-IR data from \IRAS\ and \ISO, although \citet{klaas2001} describe
the existing {\it ISO\/} photometry as ``partly distorted and
uncertain''.  \citet{benford1999} reports a measurement at
350\,\micron\ of $1.05\pm0.25\,$Jy, which is consistent with the BLAST
350\,\micron\ photometry.  There is also 450\,\micron\ photometry
published, although using a much smaller aperture
\citep{rigopoulou1996}.  BLAST flux densities thus help constrain the
SED in the poorly-sampled submillimeter regime (see
Figure~\ref{sed:mrk231}).  However, given the relative faintness of
this source, the improvement is modest compared with other sources
reported here.  We find slightly lower values for temperature and
$\beta$ compared with \citet{klaas2001}, but they also claim that the data
are better fit with multiple temperature components.

\begin{figure*}
\includegraphics[width=2in]{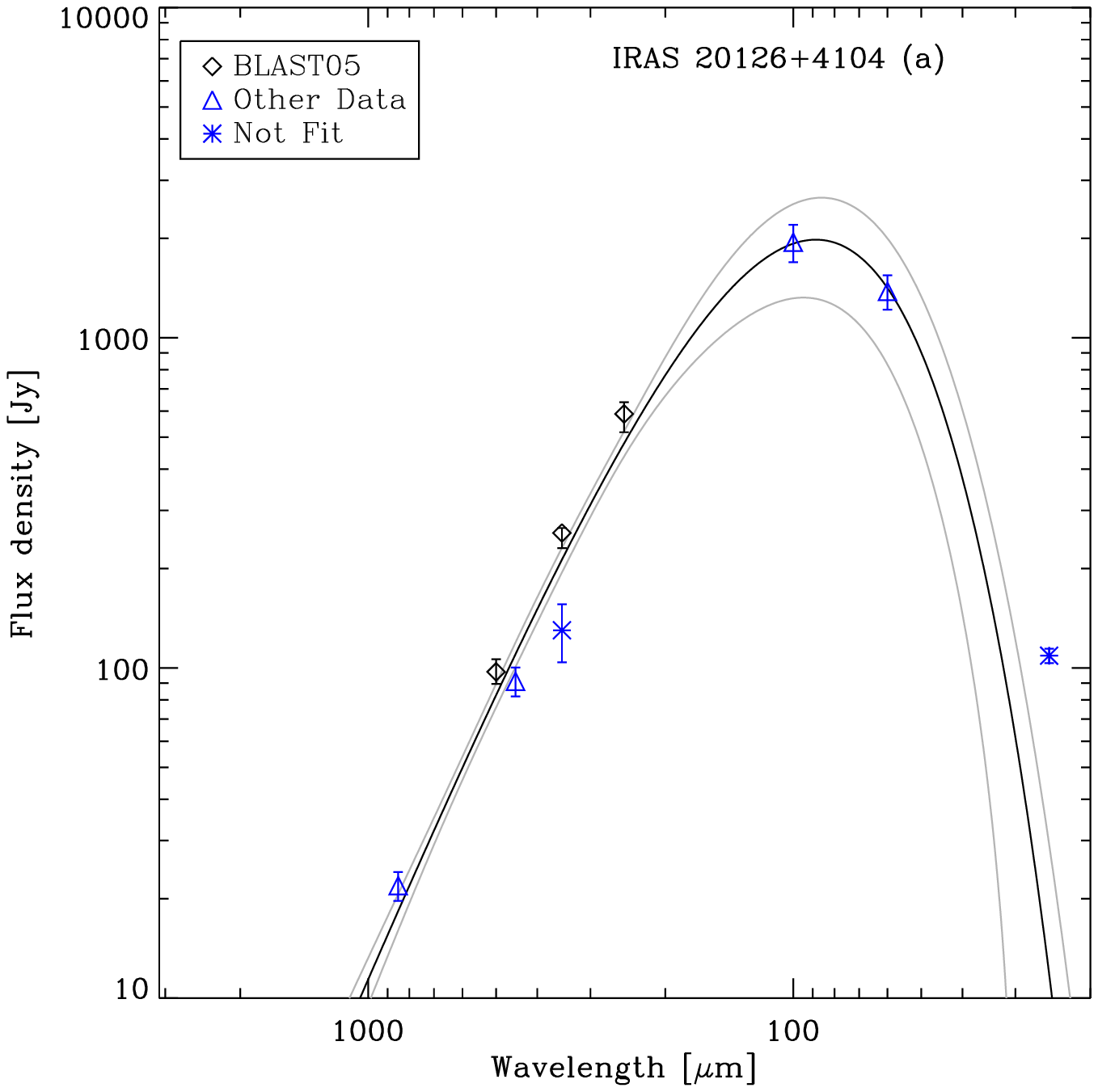}
\includegraphics[width=2in]{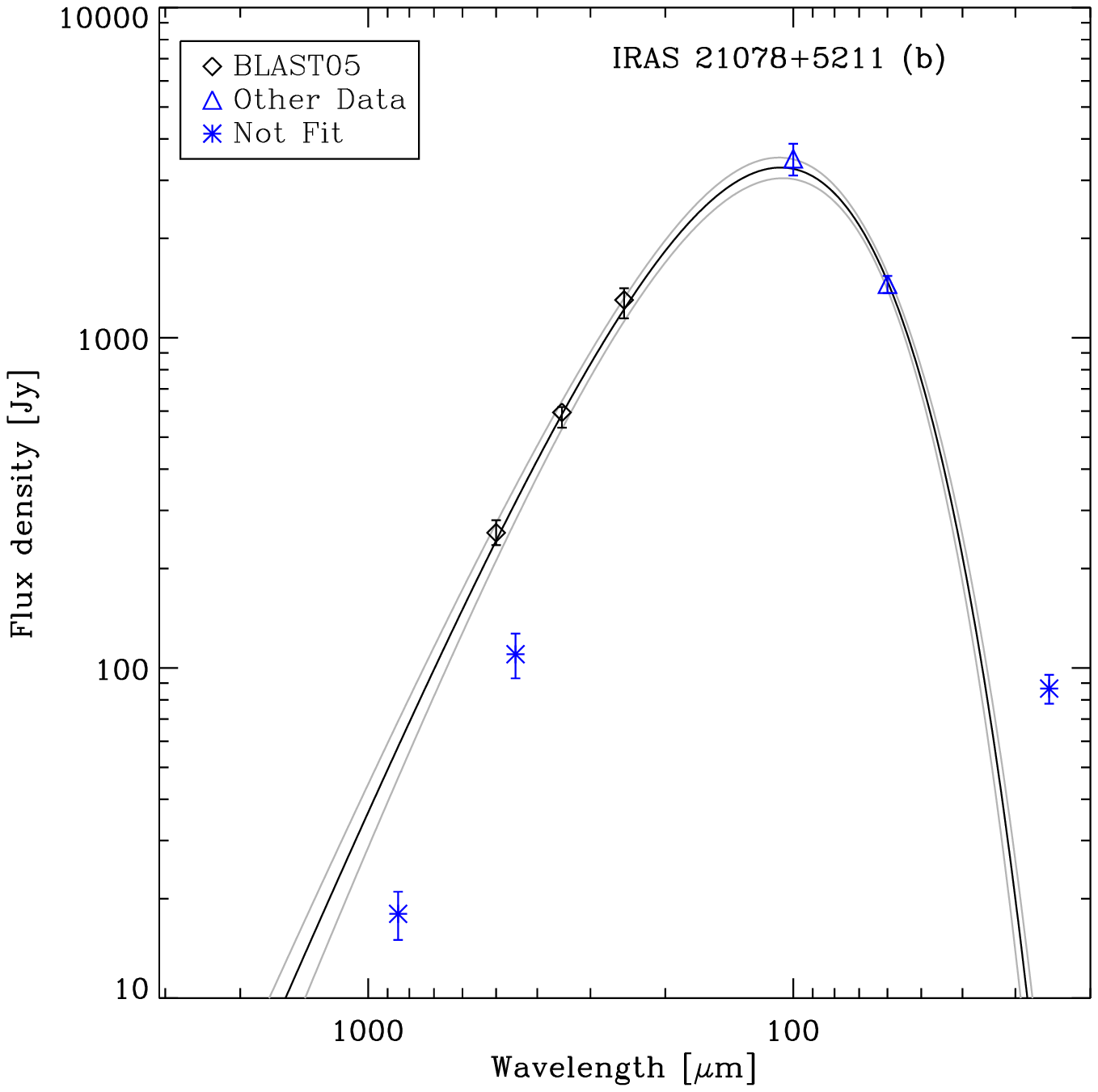}
\includegraphics[width=2in]{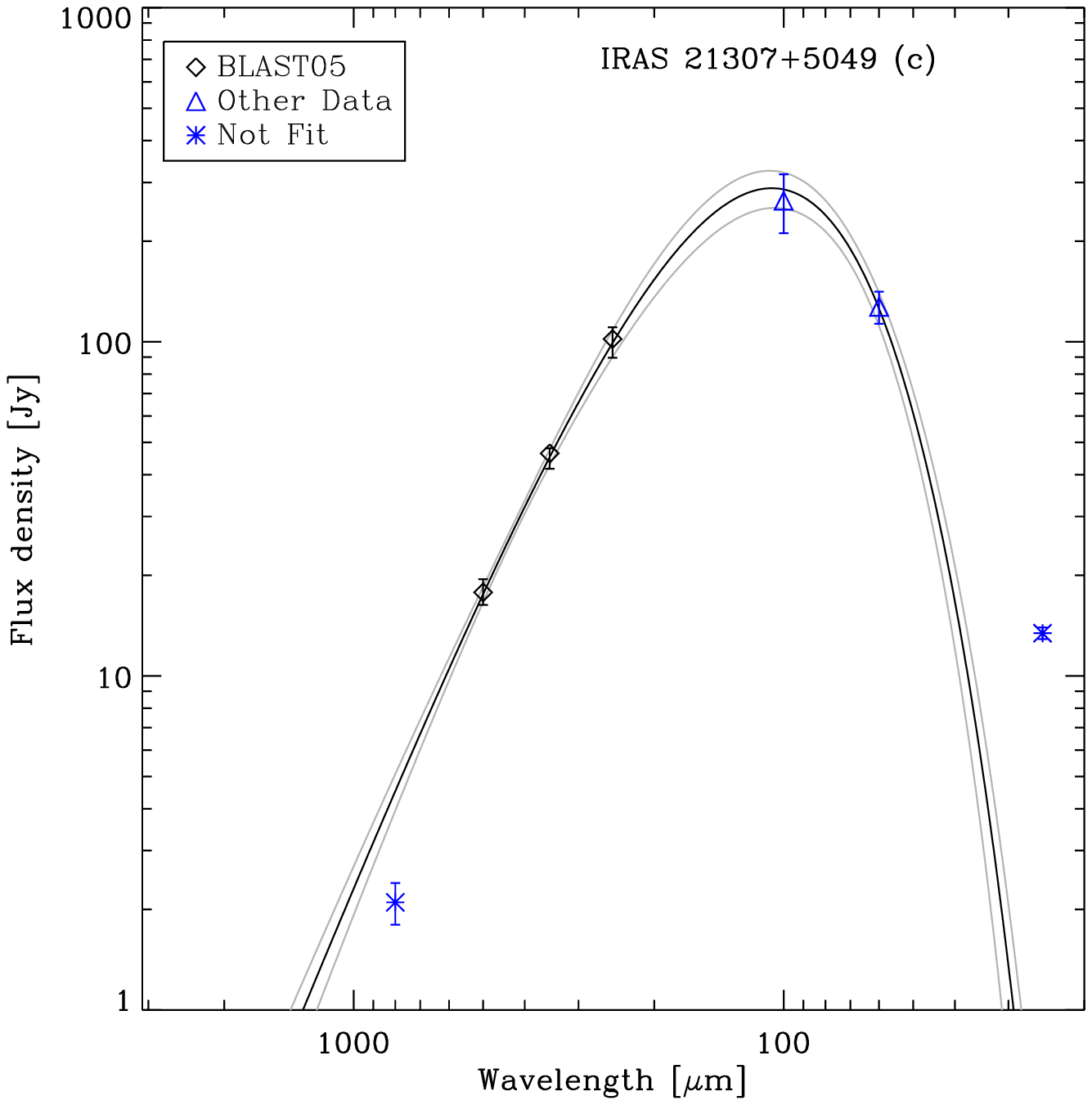}
\includegraphics[width=2in]{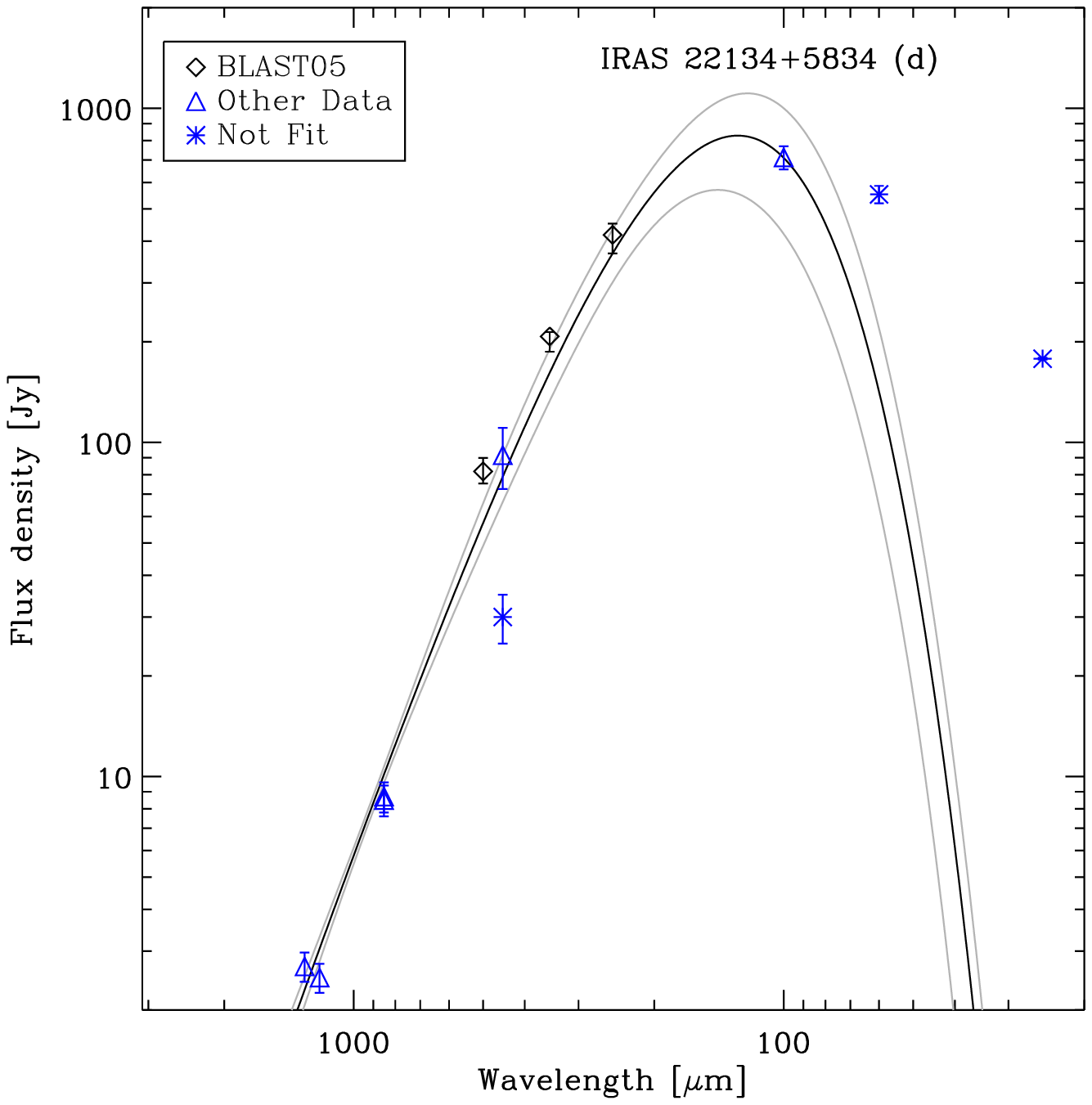}
\includegraphics[width=2in]{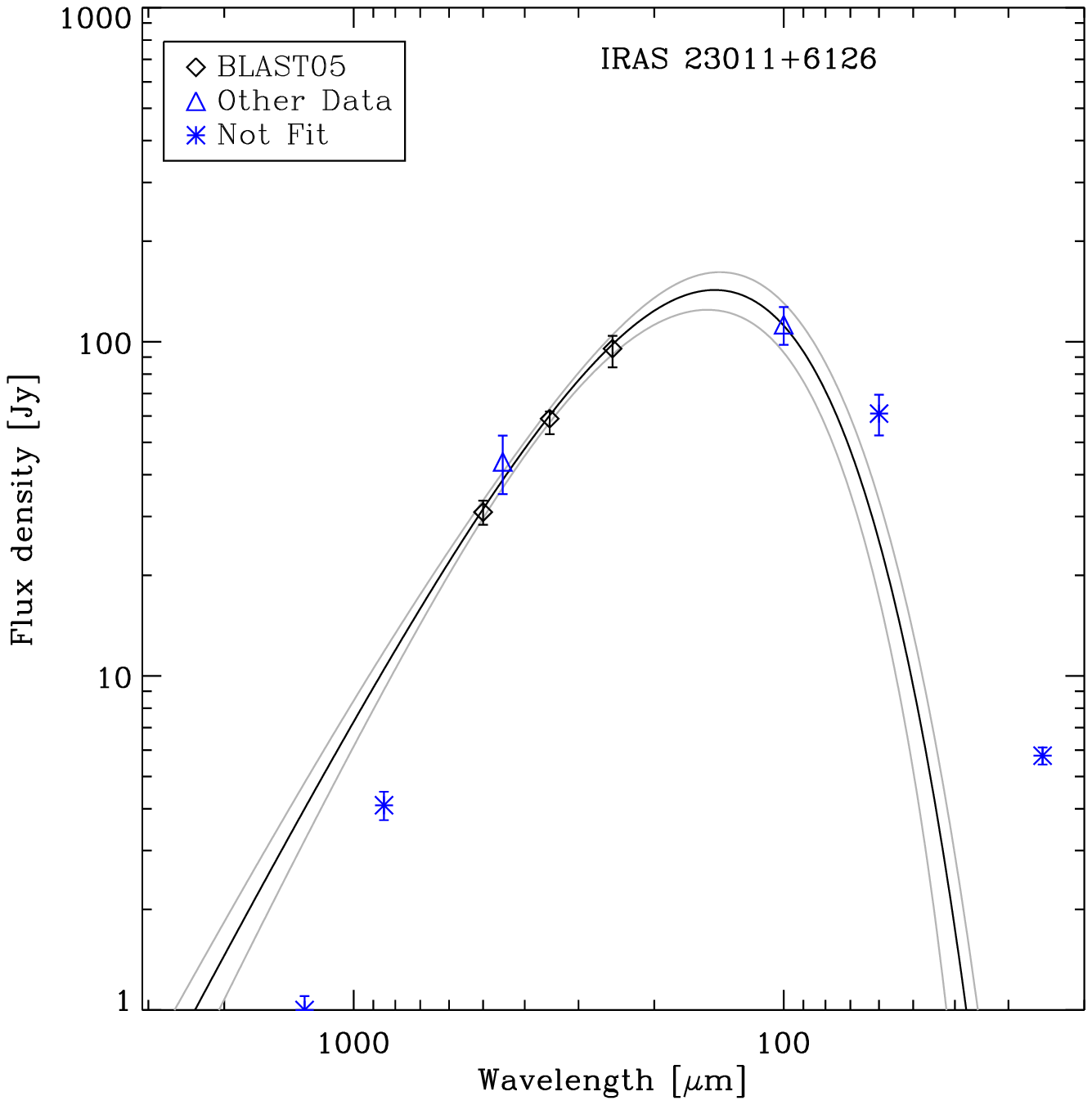}
\caption{Spectral energy distributions of compact protostellar
sources, including the BLAST fluxes (see Table \ref{calib}).  The data symbols and
fitted curves have the same meaning as those described in
Fig. \ref{sed:l1014} -- (a) IRAS~20126+4104 --- Additional data are
taken from \citet{helou1988}, \citet{hunter2000}, \citet{beuther2002},
and \citet{williams2004}.  The
350\,\micron\ data point from \citet{hunter2000} is not used in the
fit, due to the smaller aperture.  -- (b) IRAS~21078+5211 --- Additional
data are taken from \citet{helou1988} and \citet{jenness1995}.  Three
data points are not used in the fit due to
the small effective aperture sizes in those measurements.  -- (c)
IRAS~21307+5049 --- Additional data are taken from \citet{molinari2000}
and \citet{helou1988}.  The 850\,\micron\ data point from
\citet{molinari2000} is not used in the fit due to its smaller
aperture compared to the BLAST PSF.  -- (d) IRAS~22134+5834 ---
Additional data are taken from \citet{helou1988}, 
\citet{chini2001}, \citet{beuther2002}, and \citet{williams2004}.  
-- (e) IRAS~23011+6126 --- Additional data are taken
from \citet{helou1988} and \citet{chini2001}.  The 850\,\micron\ data
point from \citet{chini2001} is not used in the fit due to the smaller
aperture.  }
\label{sed:protostellar}
\end{figure*}

\subsection{Compact Protostellar Sources} \label{sec:sources}

The remaining Galactic targets, described below, are all compact
protostellar sources, identified via colour-criteria in the FIR, that
satisfy the BLAST selection requirements. 
We present here only their fluxes,
dust masses and luminosities (Tables 2 \& 3) derived from the
best-fit models describing their SEDs (Fig~\ref{sed:protostellar}).

\subsubsection{IRAS~20126+4104}
IRAS~20126+4104 is a high-mass compact protostar, located towards the
Cygnus~X molecular cloud complex, although its exact distance is
uncertain. We adopt a distance of $1.7\,$kpc \citep{shepherd2000}.  It
is probably the best studied example of a massive proto-stellar system
associated with a disk and outflow \citep{edris2005}. 
BLAST
provides better constraints on the SED \citep{wilking1989,hunter2000},
particularly the submillimeter slope, as shown in
Figure~\ref{sed:protostellar}. 

\subsubsection{IRAS~21078+5211}
IRAS~21078+5211 is a compact \ion{H}{2} region with a water maser,
located at a distance of approximately 1.65\,kpc
\citep{molinari1996,wouterloot1993}.
The best-fit
single-temperature model with $T\,{\simeq}\,34\,$K and $\beta\,{\simeq}\,1.1$,
is consistent with a previous study by ~\cite{mccutcheon1991}.

\subsubsection{IRAS~21307+5049}
IRAS~21307+5049 \citep[also Mol~136,][]{molinari1996} is a high mass
protostellar candidate.  At a distance of 3.6\,kpc, it has an
angular size of about 4\arcsec\ in the millimeter \citep{molinari2002}, 
and is therefore point-like
to BLAST\@. \citet{fontani2004} present 850\micron\ SCUBA continuum
observations, detecting an extended halo of 40\arcsec\ diameter and a
compact unresolved core. The extended halo is not detected by BLAST\@.
We only fit a
single-temperature modified blackbody curve to the cold core
component.  We find a temperature of $32\,$K and luminosity of $L_{\rm
FIR}\,{\simeq}\,4{,}000\,{\rm L}_\odot$, in very good agreement with
\citet{molinari2000}.

\subsubsection{IRAS~22134+5834}
IRAS~22134+5834 is a medium-to-high mass protostar located within the
extensive \ion{H}{2} region S134, in the constellation
Cepheus.  It is at a distance of approximately 900\,pc \citep{dobashi1994}, and
has a total luminosity of ${\sim}\,1600\,{\rm L}_{\odot}$\@.

\subsubsection{IRAS~23011+6126}
IRAS~23011+6126 is a young protostar located at a distance of
$\sim 730$\,pc \citep{crawford1970}.  IRAS~23011+6126 is point-like to
BLAST\@.  \citet{chini2001} discusses the current best SED (similar to
IRAS~22134+5834 above).  The peak of the SED is at about 150\,\micron,
as shown in Figure~\ref{sed:protostellar}.  It appears to have
$T\,{\simeq}\,28\,$K, with a shallower slope ($\beta = 0.6 \pm 0.2$)
than most of the other protostellar sources that BLAST observed.

\section{Conclusions}
For the BLAST 2005 flight, the out-of-focus PSFs significantly reduced the
point-source sensitivity, and yet because this is such a difficult waveband to
observe from the ground, the BLAST05 data still produced useful results.  The
submillimeter flux densities reported here agree well with other measurements,
at least when the aperture sizes are consistent.  However, BLAST05 uncertainties are
typically much smaller than those of other measurements in the
250--500\,\micron\ regime. 

A particular advantage of the BLAST data is the ability to constrain the
emissivity index $\beta$, due primarily to the 3 separate submillimeter
channels, observed simultaneously with the same telescope, and having a
consistent calibration procedure performed across all 3 bands.  

Arp~220 was adopted as the primary calibrator for the BLAST05 data.  A
useful cross-check is to note that the derived flux densities for
Pallas were all within $1\sigma$ of the values derived from the TPM
\citep{muller2002}. 

By applying this careful calibration procedure, BLAST has been able to
improve estimates of dust temperatures and masses for several
relatively bright sources, and also to provide a database of
submillimeter measurements which may be useful for calibrating future
instruments. 

\acknowledgments
The BLAST collaboration acknowledges the support of
NASA through grant numbers NAG5-12785, NAG5-13301, and NNGO-6GI11G, the
Canadian Space Agency (CSA), Canada's Natural Sciences and Engineering
Research Council (NSERC), and the UK Particle Physics \& Astronomy
Research Council (PPARC).
We would also like to thank the Columbia Scientific Balloon Facility (CSBF)
staff for their outstanding work, as well as
T.G.~M{\"u}ller for valuable discussions on Pallas.
LO acknowledges partial support by the Puerto Rico
Space Grant Consortium and by the Fondo Istitucional 
para la Investigacion of the University of Puerto Rico.
CBN acknowledges support from the Canadian Institute for Advanced Research.
This research has been enabled by the use of WestGrid computing resources.
This research also made use of the SIMBAD database, operated at the
Centre de Don{\'e}es astronomiques de Strasbourg (CDS), Strasbourg, France,
and the NASA/IPAC Extragalactic Database (NED), which is operated by the
Jet Propulsion Laboratory, California Institute of
Technology, under contract with the National
Aeronautics and Space Administration.

\bibliographystyle{apj}
\bibliography{ms}

\begin{thebibliography}{83}
\expandafter\ifx\csname natexlab\endcsname\relax\def\natexlab#1{#1}\fi

\bibitem[{{Aretxaga} {et~al.}(2005){Aretxaga}, {Hughes}, \&
  {Dunlop}}]{aretxaga2005}
{Aretxaga}, I., {Hughes}, D.~H., \& {Dunlop}, J.~S. 2005, \mnras, 358, 1240

\bibitem[{{Arp}(1966)}]{arp1966}
{Arp}, H. 1966, \apjs, 14, 1

\bibitem[{{Beelen} {et~al.}(2006){Beelen}, {Cox}, {Benford}, {Dowell},
  {Kov{\'a}cs}, {Bertoldi}, {Omont}, \& {Carilli}}]{beelen2006}
{Beelen}, A., {Cox}, P., {Benford}, D.~J., {Dowell}, C.~D., {Kov{\'a}cs}, A.,
  {Bertoldi}, F., {Omont}, A., \& {Carilli}, C.~L. 2006, \apj, 642, 694

\bibitem[{{Benford}(1999)}]{benford1999}
{Benford}, D.~J. 1999, PhD thesis, California Institute of Technology

\bibitem[{{Beuther} {et~al.}(2002){Beuther}, {Schilke}, {Menten}, {Motte},
  {Sridharan}, \& {Wyrowski}}]{beuther2002}
{Beuther}, H., {Schilke}, P., {Menten}, K.~M., {Motte}, F., {Sridharan}, T.~K.,
  \& {Wyrowski}, F. 2002, \apj, 566, 945

\bibitem[{{Bourke} {et~al.}(2005){Bourke}, {Crapsi}, {Myers}, {Evans},
  {Wilner}, {Huard}, {J{\o}rgensen}, \& {Young}}]{bourke2005}
{Bourke}, T.~L., {Crapsi}, A., {Myers}, P.~C., {Evans}, II, N.~J., {Wilner},
  D.~J., {Huard}, T.~L., {J{\o}rgensen}, J.~K., \& {Young}, C.~H. 2005, \apjl,
  633, L129

\bibitem[{{Campbell} {et~al.}(1982){Campbell}, {Niles}, {Nawfel}, {Hawrylycz},
  {Hoffmann}, \& {Thronson}}]{campbell1982}
{Campbell}, M.~F., {Niles}, D., {Nawfel}, R., {Hawrylycz}, M., {Hoffmann},
  W.~F., \& {Thronson}, Jr., H.~A. 1982, \apj, 261, 550

\bibitem[{{Chapin} {et~al.}(2008)}]{chapin2008}
{Chapin}, E. {et~al.} 2008, ApJ, in press

\bibitem[{{Chini} {et~al.}(2001){Chini}, {Ward-Thompson}, {Kirk}, {Nielbock},
  {Reipurth}, \& {Sievers}}]{chini2001}
{Chini}, R., {Ward-Thompson}, D., {Kirk}, J.~M., {Nielbock}, M., {Reipurth},
  B., \& {Sievers}, A. 2001, \aap, 369, 155

\bibitem[{{Clegg} {et~al.}(1976){Clegg}, {Rowan-Robinson}, \&
  {Ade}}]{clegg1976}
{Clegg}, P.~E., {Rowan-Robinson}, M., \& {Ade}, P.~A.~R. 1976, \aj, 81, 399

\bibitem[{{Cohen} \& {Kuhi}(1977)}]{cohen1977}
{Cohen}, M. \& {Kuhi}, L.~V. 1977, \apj, 213, 79

\bibitem[{{Colome} {et~al.}(1995){Colome}, {Harvey}, {Lester}, {Campbell}, \&
  {Butner}}]{colome1995}
{Colome}, C., {Harvey}, P.~M., {Lester}, D.~F., {Campbell}, M.~F., \& {Butner},
  H.~M. 1995, \apj, 447, 236

\bibitem[{{Coppin} {et~al.}(2006){Coppin}, {Chapin}, {Mortier}, {Scott},
  {Borys}, {Dunlop}, {Halpern}, {Hughes}, {Pope}, {Scott}, {Serjeant}, {Wagg},
  {Alexander}, {Almaini}, {Aretxaga}, {Babbedge}, {Best}, {Blain}, {Chapman},
  {Clements}, {Crawford}, {Dunne}, {Eales}, {Edge}, {Farrah}, {Gazta{\~n}aga},
  {Gear}, {Granato}, {Greve}, {Fox}, {Ivison}, {Jarvis}, {Jenness}, {Lacey},
  {Lepage}, {Mann}, {Marsden}, {Martinez-Sansigre}, {Oliver}, {Page},
  {Peacock}, {Pearson}, {Percival}, {Priddey}, {Rawlings}, {Rowan-Robinson},
  {Savage}, {Seigar}, {Sekiguchi}, {Silva}, {Simpson}, {Smail}, {Stevens},
  {Takagi}, {Vaccari}, {van Kampen}, \& {Willott}}]{coppin2006}
{Coppin}, K., {Chapin}, E.~L., {Mortier}, A.~M.~J., {Scott}, S.~E., {Borys},
  C., {Dunlop}, J.~S., {Halpern}, M., {Hughes}, D.~H., {Pope}, A., {Scott}, D.,
  {Serjeant}, S., {Wagg}, J., {Alexander}, D.~M., {Almaini}, O., {Aretxaga},
  I., {Babbedge}, T., {Best}, P.~N., {Blain}, A., {Chapman}, S., {Clements},
  D.~L., {Crawford}, M., {Dunne}, L., {Eales}, S.~A., {Edge}, A.~C., {Farrah},
  D., {Gazta{\~n}aga}, E., {Gear}, W.~K., {Granato}, G.~L., {Greve}, T.~R.,
  {Fox}, M., {Ivison}, R.~J., {Jarvis}, M.~J., {Jenness}, T., {Lacey}, C.,
  {Lepage}, K., {Mann}, R.~G., {Marsden}, G., {Martinez-Sansigre}, A.,
  {Oliver}, S., {Page}, M.~J., {Peacock}, J.~A., {Pearson}, C.~P., {Percival},
  W.~J., {Priddey}, R.~S., {Rawlings}, S., {Rowan-Robinson}, M., {Savage},
  R.~S., {Seigar}, M., {Sekiguchi}, K., {Silva}, L., {Simpson}, C., {Smail},
  I., {Stevens}, J.~A., {Takagi}, T., {Vaccari}, M., {van Kampen}, E., \&
  {Willott}, C.~J. 2006, \mnras, 372, 1621

\bibitem[{{Coppin} {et~al.}(2008){Coppin}, {Halpern}, {Scott}, {Borys},
  {Dunlop}, {Dunne}, {Ivison}, {Wagg}, {Aretxaga}, {Battistelli}, {Benson},
  {Blain}, {Chapman}, {Clements}, {Dye}, {Farrah}, {Hughes}, {Jenness}, {van
  Kampen}, {Lacey}, {Mortier}, {Pope}, {Priddey}, {Serjeant}, {Smail},
  {Stevens}, \& {Vaccari}}]{coppin2008}
{Coppin}, K., {Halpern}, M., {Scott}, D., {Borys}, C., {Dunlop}, J., {Dunne},
  L., {Ivison}, R., {Wagg}, J., {Aretxaga}, I., {Battistelli}, E., {Benson},
  A., {Blain}, A., {Chapman}, S., {Clements}, D., {Dye}, S., {Farrah}, D.,
  {Hughes}, D., {Jenness}, T., {van Kampen}, E., {Lacey}, C., {Mortier}, A.,
  {Pope}, A., {Priddey}, R., {Serjeant}, S., {Smail}, I., {Stevens}, J., \&
  {Vaccari}, M. 2008, \mnras, 123

\bibitem[{{Cox} {et~al.}(1996){Cox}, {Gonzalez-Alfonso}, {Barlow}, {Liu},
  {Lim}, {Swinyard}, {Cernicharo}, {Omont}, {Caux}, {Gry}, {Griffin},
  {Baluteau}, {Clegg}, {Sidher}, {Pequignot}, {Nguyen-Q-Rieu}, {King}, {Ade},
  {Towlson}, {Emery}, {Furniss}, {Joubert}, {Skinner}, {Cohen}, {Armand},
  {Burgdorf}, {Eward}, {di Giorgio}, {Molinari}, {Texier}, {Trams}, {Unger},
  {Glencross}, {Lorenzetti}, {Nisini}, {Orfei}, {Saraceno}, \&
  {Serra}}]{cox1996}
{Cox}, P., {Gonzalez-Alfonso}, E., {Barlow}, M.~J., {Liu}, X.-W., {Lim}, T.,
  {Swinyard}, B.~M., {Cernicharo}, J., {Omont}, A., {Caux}, E., {Gry}, C.,
  {Griffin}, M.~J., {Baluteau}, J.-P., {Clegg}, P.~E., {Sidher}, S.,
  {Pequignot}, D., {Nguyen-Q-Rieu}, {King}, K.~J., {Ade}, P.~A.~R., {Towlson},
  W.~A., {Emery}, R.~J., {Furniss}, I., {Joubert}, M., {Skinner}, C.~J.,
  {Cohen}, M., {Armand}, C., {Burgdorf}, M., {Eward}, D., {di Giorgio}, A.,
  {Molinari}, S., {Texier}, D., {Trams}, N., {Unger}, S.~J., {Glencross},
  W.~M., {Lorenzetti}, D., {Nisini}, B., {Orfei}, R., {Saraceno}, P., \&
  {Serra}, G. 1996, \aap, 315, L265

\bibitem[{{Crampton} {et~al.}(1975){Crampton}, {Cowley}, \&
  {Humphreys}}]{crampton1975}
{Crampton}, D., {Cowley}, A.~P., \& {Humphreys}, R.~M. 1975, \apjl, 198, L135

\bibitem[{{Crawford} \& {Barnes}(1970)}]{crawford1970}
{Crawford}, D.~L. \& {Barnes}, J.~V. 1970, \aj, 75, 952

\bibitem[{{Davis} {et~al.}(2007){Davis}, {Kumar}, {Sandell}, {Froebrich},
  {Smith}, \& {Currie}}]{davis2007}
{Davis}, C.~J., {Kumar}, M.~S.~N., {Sandell}, G., {Froebrich}, D., {Smith},
  M.~D., \& {Currie}, M.~J. 2007, \mnras, 374, 29

\bibitem[{{Do} {et~al.}(2005){Do}, {Morris}, {Sahai}, \&
  {Stapelfeldt}}]{do2005}
{Do}, T., {Morris}, M., {Sahai}, R., \& {Stapelfeldt}, K.~R. 2005, in Bulletin
  of the American Astronomical Society, 1161

\bibitem[{{Dobashi} {et~al.}(1994){Dobashi}, {Bernard}, {Yonekura}, \&
  {Fukui}}]{dobashi1994}
{Dobashi}, K., {Bernard}, J.-P., {Yonekura}, Y., \& {Fukui}, Y. 1994, \apjs,
  95, 419

\bibitem[{{Downes} \& {Rinehart}(1966)}]{downes1966}
{Downes}, D. \& {Rinehart}, R. 1966, \apj, 144, 937

\bibitem[{{Dunne} \& {Eales}(2001)}]{dunne2001}
{Dunne}, L. \& {Eales}, S.~A. 2001, \mnras, 327, 697

\bibitem[{{Edris} {et~al.}(2005){Edris}, {Fuller}, {Cohen}, \&
  {Etoka}}]{edris2005}
{Edris}, K.~A., {Fuller}, G.~A., {Cohen}, R.~J., \& {Etoka}, S. 2005, \aap,
  434, 213

\bibitem[{{Fontani} {et~al.}(2004){Fontani}, {Cesaroni}, {Testi}, {Molinari},
  {Zhang}, {Brand}, \& {Walmsley}}]{fontani2004}
{Fontani}, F., {Cesaroni}, R., {Testi}, L., {Molinari}, S., {Zhang}, Q.,
  {Brand}, J., \& {Walmsley}, C.~M. 2004, \aap, 424, 179

\bibitem[{{Friesen} {et~al.}(2005){Friesen}, {Johnstone}, {Naylor}, \&
  {Davis}}]{friesen2005}
{Friesen}, R.~K., {Johnstone}, D., {Naylor}, D.~A., \& {Davis}, G.~R. 2005,
  \mnras, 361, 460

\bibitem[{{Griffin} {et~al.}(2002){Griffin}, {Bock}, \& {Gear}}]{griffin2002}
{Griffin}, M.~J., {Bock}, J.~J., \& {Gear}, W.~K. 2002, Applied Optics, 41,
  6543

\bibitem[{{Griffin} \& {Orton}(1993)}]{griffin1993}
{Griffin}, M.~J. \& {Orton}, G.~S. 1993, Icarus, 105, 537

\bibitem[{{Griffin} {et~al.}(2004){Griffin}, {Swinyard}, \&
  {Vigroux}}]{griffin2004}
{Griffin}, M.~J., {Swinyard}, B.~M., \& {Vigroux}, L. 2004, in Optical,
  Infrared, and Millimeter Space Telescopes. Edited by Mather, John C.
  Proceedings of the SPIE, Volume 5487, ed. J.~C. {Mather}, 413--424

\bibitem[{{Hargrave} {et~al.}(2006){Hargrave}, {Waskett}, {Lim}, \&
  {Swinyard}}]{hargrave2006}
{Hargrave}, P., {Waskett}, T., {Lim}, T., \& {Swinyard}, B. 2006, in Millimeter
  and Submillimeter Detectors and Instrumentation for Astronomy III. Edited by
  Zmuidzinas, Jonas; Holland, Wayne S.; Withington, Stafford; Duncan, William
  D.. Proceedings of the SPIE, Volume 6275, pp. 627514 (2006).

\bibitem[{{Hargrave} {et~al.}(2008)}]{hargrave2008}
{Hargrave}, P. {et~al.} 2008, ApJ, submitted

\bibitem[{{Harvey} {et~al.}(1977){Harvey}, {Campbell}, \&
  {Hoffmann}}]{harvey1977}
{Harvey}, P.~M., {Campbell}, M.~F., \& {Hoffmann}, W.~F. 1977, \apj, 211, 786

\bibitem[{{Helou} \& {Walker}(1988)}]{helou1988}
{Helou}, G. \& {Walker}, D.~W., eds. 1988, {Infrared astronomical satellite
  ({\it IRAS}) catalogs and atlases}

\bibitem[{{Howard} {et~al.}(1996){Howard}, {Pipher}, {Forrest}, \& {de
  Pree}}]{howard1996}
{Howard}, E.~M., {Pipher}, J.~L., {Forrest}, W.~J., \& {de Pree}, C.~G. 1996,
  \apj, 460, 744

\bibitem[{{Huard} {et~al.}(2006){Huard}, {Myers}, {Murphy}, {Crews}, {Lada},
  {Bourke}, {Crapsi}, {Evans}, {McCarthy}, \& {Kulesa}}]{huard2006}
{Huard}, T.~L., {Myers}, P.~C., {Murphy}, D.~C., {Crews}, L.~J., {Lada}, C.~J.,
  {Bourke}, T.~L., {Crapsi}, A., {Evans}, II, N.~J., {McCarthy}, Jr., D.~W., \&
  {Kulesa}, C. 2006, \apj, 640, 391

\bibitem[{{Hunter} {et~al.}(2000){Hunter}, {Churchwell}, {Watson}, {Cox},
  {Benford}, \& {Roelfsema}}]{hunter2000}
{Hunter}, T.~R., {Churchwell}, E., {Watson}, C., {Cox}, P., {Benford}, D.~J.,
  \& {Roelfsema}, P.~R. 2000, \aj, 119, 2711

\bibitem[{{Jakob} {et~al.}(2007){Jakob}, {Kramer}, {Simon}, {Schneider},
  {Ossenkopf}, {Bontemps}, {Graf}, \& {Stutzki}}]{jakob2007}
{Jakob}, H., {Kramer}, C., {Simon}, R., {Schneider}, N., {Ossenkopf}, V.,
  {Bontemps}, S., {Graf}, U.~U., \& {Stutzki}, J. 2007, \aap, 461, 999

\bibitem[{{Jenness} {et~al.}(1995){Jenness}, {Scott}, \&
  {Padman}}]{jenness1995}
{Jenness}, T., {Scott}, P.~F., \& {Padman}, R. 1995, \mnras, 276, 1024

\bibitem[{{Jenness} {et~al.}(2002){Jenness}, {Stevens}, {Archibald},
  {Economou}, {Jessop}, \& {Robson}}]{jenness2002}
{Jenness}, T., {Stevens}, J.~A., {Archibald}, E.~N., {Economou}, F., {Jessop},
  N.~E., \& {Robson}, E.~I. 2002, \mnras, 336, 14

\bibitem[{{Johnstone} {et~al.}(2003){Johnstone}, {Boonman}, \& {van
  Dishoeck}}]{johnstone2003}
{Johnstone}, D., {Boonman}, A.~M.~S., \& {van Dishoeck}, E.~F. 2003, \aap, 412,
  157

\bibitem[{{Klaas} {et~al.}(2001){Klaas}, {Haas}, {M{\"u}ller}, {Chini},
  {Schulz}, {Coulson}, {Hippelein}, {Wilke}, {Albrecht}, \&
  {Lemke}}]{klaas2001}
{Klaas}, U., {Haas}, M., {M{\"u}ller}, S.~A.~H., {Chini}, R., {Schulz}, B.,
  {Coulson}, I., {Hippelein}, H., {Wilke}, K., {Albrecht}, M., \& {Lemke}, D.
  2001, \aap, 379, 823

\bibitem[{{Kumar} {et~al.}(2007){Kumar}, {Davis}, {Grave}, {Ferreira}, \&
  {Froebrich}}]{kumar2007}
{Kumar}, M.~S.~N., {Davis}, C.~J., {Grave}, J.~M.~C., {Ferreira}, B., \&
  {Froebrich}, D. 2007, \mnras, 374, 54

\bibitem[{{L{\'{\i}}pari} {et~al.}(2005){L{\'{\i}}pari}, {Terlevich}, {Zheng},
  {Garcia-Lorenzo}, {Sanchez}, \& {Bergmann}}]{lipari2005}
{L{\'{\i}}pari}, S., {Terlevich}, R., {Zheng}, W., {Garcia-Lorenzo}, B.,
  {Sanchez}, S.~F., \& {Bergmann}, M. 2005, \mnras, 360, 416

\bibitem[{{Lisenfeld} {et~al.}(2000){Lisenfeld}, {Isaak}, \&
  {Hills}}]{lisenfeld2000}
{Lisenfeld}, U., {Isaak}, K.~G., \& {Hills}, R. 2000, \mnras, 312, 433

\bibitem[{{Lynds}(1962)}]{lynds1962}
{Lynds}, B.~T. 1962, \apjs, 7, 1

\bibitem[{{Markarian}(1969)}]{markarian1969}
{Markarian}, B.~E. 1969, Astrofizika, 5, 286

\bibitem[{{McCutcheon} {et~al.}(1991){McCutcheon}, {Sato}, {Dewdney}, \&
  {Purton}}]{mccutcheon1991}
{McCutcheon}, W.~H., {Sato}, T., {Dewdney}, P.~E., \& {Purton}, C.~R. 1991,
  \aj, 101, 1435

\bibitem[{{Mitchell} {et~al.}(1996){Mitchell}, {Ostro}, {Hudson}, {Rosema},
  {Campbell}, {Velez}, {Chandler}, {Shapiro}, {Giorgini}, \&
  {Yeomans}}]{mitchell1996}
{Mitchell}, D.~L., {Ostro}, S.~J., {Hudson}, R.~S., {Rosema}, K.~D.,
  {Campbell}, D.~B., {Velez}, R., {Chandler}, J.~F., {Shapiro}, I.~I.,
  {Giorgini}, J.~D., \& {Yeomans}, D.~K. 1996, Icarus, 124, 113

\bibitem[{{Molinari} {et~al.}(1996){Molinari}, {Brand}, {Cesaroni}, \&
  {Palla}}]{molinari1996}
{Molinari}, S., {Brand}, J., {Cesaroni}, R., \& {Palla}, F. 1996, \aap, 308,
  573

\bibitem[{{Molinari} {et~al.}(2000){Molinari}, {Brand}, {Cesaroni}, \&
  {Palla}}]{molinari2000}
---. 2000, \aap, 355, 617

\bibitem[{{Molinari} {et~al.}(2002){Molinari}, {Testi}, {Rodr{\'{\i}}guez}, \&
  {Zhang}}]{molinari2002}
{Molinari}, S., {Testi}, L., {Rodr{\'{\i}}guez}, L.~F., \& {Zhang}, Q. 2002,
  \apj, 570, 758

\bibitem[{{Morita} {et~al.}(2006){Morita}, {Watanabe}, {Sugitani}, {Itoh},
  {Uehara}, {Nagashima}, {Ebizuka}, {Hasegawa}, {Kinugasa}, \&
  {Tamura}}]{morita2006}
{Morita}, A., {Watanabe}, M., {Sugitani}, K., {Itoh}, Y., {Uehara}, M.,
  {Nagashima}, C., {Ebizuka}, N., {Hasegawa}, T., {Kinugasa}, K., \& {Tamura},
  M. 2006, \pasj, 58, L41

\bibitem[{M\"{u}ller(2005)}]{muller2005}
M\"{u}ller, T.~G. 2005, Private Communication

\bibitem[{{M{\"u}ller} \& {Lagerros}(2002)}]{muller2002}
{M{\"u}ller}, T.~G. \& {Lagerros}, J.~S.~V. 2002, \aap, 381, 324

\bibitem[{{Omont} {et~al.}(1995){Omont}, {Moseley}, {Cox}, {Glaccum}, {Casey},
  {Forveille}, {Chan}, {Szczerba}, {Loewenstein}, {Harvey}, \&
  {Kwok}}]{omont1995}
{Omont}, A., {Moseley}, S.~H., {Cox}, P., {Glaccum}, W., {Casey}, S.,
  {Forveille}, T., {Chan}, K.-W., {Szczerba}, R., {Loewenstein}, R.~F.,
  {Harvey}, P.~M., \& {Kwok}, S. 1995, \apj, 454, 819

\bibitem[{{Pantachon} {et~al.}(2008)}]{pantachon2008}
{Pantachon}, G. {et~al.} 2008, ApJ, in press

\bibitem[{{Pascale} {et~al.}(2008)}]{pascale2008}
{Pascale}, E. {et~al.} 2008, ApJ, in press

\bibitem[{{Peeters} {et~al.}(2002){Peeters}, {Mart{\'{\i}}n-Hern{\'a}ndez},
  {Damour}, {Cox}, {Roelfsema}, {Baluteau}, {Tielens}, {Churchwell}, {Kessler},
  {Mathis}, {Morisset}, \& {Schaerer}}]{peeters2002}
{Peeters}, E., {Mart{\'{\i}}n-Hern{\'a}ndez}, N.~L., {Damour}, F., {Cox}, P.,
  {Roelfsema}, P.~R., {Baluteau}, J.-P., {Tielens}, A.~G.~G.~M., {Churchwell},
  E., {Kessler}, M.~F., {Mathis}, J.~S., {Morisset}, C., \& {Schaerer}, D.
  2002, \aap, 381, 571

\bibitem[{{Pipenbrink} \& {Wendker}(1988)}]{pipenbrink1988}
{Pipenbrink}, A. \& {Wendker}, H.~J. 1988, \aap, 191, 313

\bibitem[{{Price} \& {Murdock}(1983)}]{price1983}
{Price}, S.~D. \& {Murdock}, T.~L. 1983, AFGL-TR-0208 Environemental Research
  papers, 161, 1

\bibitem[{{Rigopoulou} {et~al.}(1996){Rigopoulou}, {Lawrence}, \&
  {Rowan-Robinson}}]{rigopoulou1996}
{Rigopoulou}, D., {Lawrence}, A., \& {Rowan-Robinson}, M. 1996, \mnras, 278,
  1049

\bibitem[{{Sandell}(1994)}]{sandell1994}
{Sandell}, G. 1994, \mnras, 271, 75

\bibitem[{{Sandell}(2003)}]{sandell2003}
{Sandell}, G. 2003, in ESA SP-481: The Calibration Legacy of the ISO Mission,
  ed. L.~{Metcalfe}, A.~{Salama}, S.~B. {Peschke}, \& M.~F. {Kessler}, 439--442

\bibitem[{{Sanders} {et~al.}(2003){Sanders}, {Mazzarella}, {Kim}, {Surace}, \&
  {Soifer}}]{sanders2003}
{Sanders}, D.~B., {Mazzarella}, J.~M., {Kim}, D.-C., {Surace}, J.~A., \&
  {Soifer}, B.~T. 2003, \aj, 126, 1607

\bibitem[{{Schlegel} {et~al.}(1998){Schlegel}, {Finkbeiner}, \&
  {Davis}}]{schlegel1998}
{Schlegel}, D.~J., {Finkbeiner}, D.~P., \& {Davis}, M. 1998, \apj, 500, 525

\bibitem[{{Scott} {et~al.}(2006){Scott}, {Chapin}, {Aretxaga}, {Austermann},
  {Coppin}, {Crowe}, {Frey}, {Gibb}, {Halpern}, {Hughes}, {Kang}, {Kim},
  {Lowenthal}, {Perera}, {Pope}, {Scott}, {Wilson}, \& {Yun}}]{scott2006}
{Scott}, D., {Chapin}, E., {Aretxaga}, I., {Austermann}, J., {Coppin}, K.,
  {Crowe}, M., {Frey}, L., {Gibb}, A., {Halpern}, M., {Hughes}, D., {Kang}, Y.,
  {Kim}, S., {Lowenthal}, J., {Perera}, T., {Pope}, A., {Scott}, K., {Wilson},
  G., \& {Yun}, M. 2006, in American Astronomical Society Meeting Abstracts,
  125.04

\bibitem[{{Scoville} {et~al.}(1997){Scoville}, {Yun}, \&
  {Bryant}}]{scoville1997}
{Scoville}, N.~Z., {Yun}, M.~S., \& {Bryant}, P.~M. 1997, \apj, 484, 702

\bibitem[{{Shepherd} {et~al.}(2000){Shepherd}, {Yu}, {Bally}, \&
  {Testi}}]{shepherd2000}
{Shepherd}, D.~S., {Yu}, K.~C., {Bally}, J., \& {Testi}, L. 2000, \apj, 535,
  833

\bibitem[{{Shirley} {et~al.}(2007){Shirley}, {Claussen}, {Bourke}, {Young}, \&
  {Blake}}]{shirley2007}
{Shirley}, Y.~L., {Claussen}, M.~J., {Bourke}, T.~L., {Young}, C.~H., \&
  {Blake}, G.~A. 2007, \apj, 667, 329

\bibitem[{{Speck} {et~al.}(2000){Speck}, {Meixner}, \& {Knapp}}]{speck2000}
{Speck}, A.~K., {Meixner}, M., \& {Knapp}, G.~R. 2000, \apjl, 545, L145

\bibitem[{{Spinoglio} {et~al.}(2002){Spinoglio}, {Andreani}, \&
  {Malkan}}]{spinoglio2002}
{Spinoglio}, L., {Andreani}, P., \& {Malkan}, M.~A. 2002, \apj, 572, 105

\bibitem[{{Stickel} {et~al.}(2004){Stickel}, {Lemke}, {Klaas}, {Krause}, \&
  {Egner}}]{stickel2004}
{Stickel}, M., {Lemke}, D., {Klaas}, U., {Krause}, O., \& {Egner}, S. 2004,
  \aap, 422, 39

\bibitem[{{Thompson} {et~al.}(2006){Thompson}, {Hatchell}, {Walsh},
  {MacDonald}, \& {Millar}}]{thompson2006}
{Thompson}, M.~A., {Hatchell}, J., {Walsh}, A.~J., {MacDonald}, G.~H., \&
  {Millar}, T.~J. 2006, \aap, 453, 1003

\bibitem[{{Thronson} \& {Harper}(1979)}]{thronson1979}
{Thronson}, Jr., H.~A. \& {Harper}, D.~A. 1979, \apj, 230, 133

\bibitem[{{Turner} {et~al.}(2001){Turner}, {Bock}, {Beeman}, {Glenn},
  {Hargrave}, {Hristov}, {Nguyen}, {Rahman}, {Sethuraman}, \& L.}]{turner2001}
{Turner}, A.~D., {Bock}, J.~J., {Beeman}, J.~W., {Glenn}, J., {Hargrave},
  P.~C., {Hristov}, V.~V., {Nguyen}, H.~T., {Rahman}, F., {Sethuraman}, S., \&
  L., W.~A. 2001, Appl.\ Opt., 40, 4921

\bibitem[{{Visser} {et~al.}(2002){Visser}, {Richer}, \&
  {Chandler}}]{visser2002}
{Visser}, A.~E., {Richer}, J.~S., \& {Chandler}, C.~J. 2002, \aj, 124, 2756

\bibitem[{{Westerhout}(1958)}]{westerhout1958}
{Westerhout}, G. 1958, \bain, 14, 215

\bibitem[{{Wilking} {et~al.}(1989){Wilking}, {Blackwell}, {Mundy}, \&
  {Howe}}]{wilking1989}
{Wilking}, B.~A., {Blackwell}, J.~H., {Mundy}, L.~G., \& {Howe}, J.~E. 1989,
  \apj, 345, 257

\bibitem[{{Williams} {et~al.}(2004){Williams}, {Fuller}, \&
  {Sridharan}}]{williams2004}
{Williams}, S.~J., {Fuller}, G.~A., \& {Sridharan}, T.~K. 2004, \aap, 417, 115

\bibitem[{{Wouterloot} {et~al.}(1993){Wouterloot}, {Brand}, \&
  {Fiegle}}]{wouterloot1993}
{Wouterloot}, J.~G.~A., {Brand}, J., \& {Fiegle}, K. 1993, \aaps, 98, 589

\bibitem[{{Wright}(2007)}]{wright2007}
{Wright}, E.~L. 2007, arXiv:astro-ph/0703640

\bibitem[{{Wynn-Williams} {et~al.}(1977){Wynn-Williams}, {Matthews}, {Werner},
  {Becklin}, \& {Neugebauer}}]{wynn-williams1977}
{Wynn-Williams}, C.~G., {Matthews}, K., {Werner}, M.~W., {Becklin}, E.~E., \&
  {Neugebauer}, G. 1977, \mnras, 179, 255

\bibitem[{{Yao} {et~al.}(2003){Yao}, {Seaquist}, {Kuno}, \& {Dunne}}]{Yao2003}
{Yao}, L., {Seaquist}, E.~R., {Kuno}, N., \& {Dunne}, L. 2003, \apj, 588, 771

\bibitem[{{Young} {et~al.}(2004){Young}, {J{\o}rgensen}, {Shirley},
  {Kauffmann}, {Huard}, {Lai}, {Lee}, {Crapsi}, {Bourke}, {Dullemond},
  {Brooke}, {Porras}, {Spiesman}, {Allen}, {Blake}, {Evans}, {Harvey},
  {Koerner}, {Mundy}, {Myers}, {Padgett}, {Sargent}, {Stapelfeldt}, {van
  Dishoeck}, {Bertoldi}, {Chapman}, {Cieza}, {DeVries}, {Ridge}, \&
  {Wahhaj}}]{young2004}
{Young}, C.~H., {J{\o}rgensen}, J.~K., {Shirley}, Y.~L., {Kauffmann}, J.,
  {Huard}, T., {Lai}, S.-P., {Lee}, C.~W., {Crapsi}, A., {Bourke}, T.~L.,
  {Dullemond}, C.~P., {Brooke}, T.~Y., {Porras}, A., {Spiesman}, W., {Allen},
  L.~E., {Blake}, G.~A., {Evans}, II, N.~J., {Harvey}, P.~M., {Koerner}, D.~W.,
  {Mundy}, L.~G., {Myers}, P.~C., {Padgett}, D.~L., {Sargent}, A.~I.,
  {Stapelfeldt}, K.~R., {van Dishoeck}, E.~F., {Bertoldi}, F., {Chapman}, N.,
  {Cieza}, L., {DeVries}, C.~H., {Ridge}, N.~A., \& {Wahhaj}, Z. 2004, \apjs,
  154, 396

\end{thebibliography}

\clearpage

\begin{deluxetable}{cccccc}
\tablecaption{Calibration Coefficients and Uncertainties for BLAST05\label{calib}}
\tablewidth{0pt}
\tablehead{
\colhead{Band} & \colhead{calib.~coeff.} & \colhead{uncertainty} &
\multicolumn{3}{c}{Pearson correlation matrix}  \\
\colhead{[\micron]} & \colhead{[$\times 10^{12}$ Jy V$^{-1}$]} & \colhead{[\%]} &
\colhead{250\,\micron} & \colhead{350\,\micron} & \colhead{500\,\micron}
}
\startdata
250   &  7.61 &   12   &   1   &  0.97  &  0.87 \\
350   &  3.16 &   10   &       &  1     &  0.96 \\
500   &  1.56 &    8   &       &        &  1    \\
\enddata
\tablecomments{Calibration coefficients, calibration uncertainties, and Pearson
correlation matrix, showing the relationship between errors in different bands
for BLAST05.}
\end{deluxetable}

\begin{deluxetable}{lccrrr}
\tabletypesize{\scriptsize}
\tablecaption{Flux Densities of BLAST05 Targeted Sources\label{fluxen}}
\tablewidth{0pt}
\tablehead{
 & \colhead{RA\tablenotemark{a}} & \colhead{DEC\tablenotemark{a}} &
\multicolumn{3}{c}{Flux Density [Jy]} \\
\colhead{Name} & \colhead{[J2000]} & \colhead{[J2000]} &
\colhead{250\,\micron} & \colhead{350\,\micron} & \colhead{500\,\micron}
}
\startdata
Pallas\tablenotemark{b}          & \nodata             & \nodata       &$11.6\pm1.5$&$ 6.3\pm0.7$&$ 3.7\pm0.3 $\\
CRL~2688        & $21\h 02\m 18\fs75$ & $+36\degr 41\arcmin 37\farcs8$ &$ 113\pm14 $&$  49\pm5  $&$21.6\pm1.7 $\\
LDN~1014        & $21\h 24\m 06\s$    & $+49\degr 59\farcm1 $          &$22.4\pm2.7$&$16.9\pm1.7$&$ 8.9\pm0.7 $\\
IRAS~20126+4104 & $20\h 14\m 25\fs1$  & $+41\degr 13\arcmin 32\arcsec$ &$ 590\pm71 $&$ 256\pm26 $&$  97\pm8   $\\
IRAS~21078+5211 & $21\h 09\m 25\fs2$  & $+52\degr 23\arcmin 44\arcsec$ &$1300\pm160$&$ 590\pm60 $&$ 260\pm21  $\\
IRAS~21307+5049 & $21\h 32\m 31\fs5$  & $+51\degr 02\arcmin 22\arcsec$ &$ 102\pm12 $&$  46\pm5  $&$  18\pm1.5 $\\
IRAS~22134+5834 & $22\h 15\m 09\fs1$  & $+58\degr 49\arcmin 09\arcsec$ &$ 418\pm50 $&$ 208\pm21 $&$  82\pm7   $\\
IRAS~23011+6126 & $23\h 03\m 13\fs9$  & $+61\degr 42\arcmin 21\arcsec$ &$  95\pm12 $&$  59\pm6  $&$  31\pm2.5 $\\
K3-50A\tablenotemark{c} & $20\h 01\m 45\fs6$  & $+33\degr 32\arcmin 42\arcsec$ &$2100\pm270$&$ 590\pm67 $&$ 270\pm26  $\\
K3-50C\tablenotemark{c} & $20\h 01\m 54\fs2$  & $+33\degr 34\arcmin 15\arcsec$ &$1870\pm240$&$1000\pm110$&$ 370\pm35  $\\
W~75N\tablenotemark{d} & $20\h 38\m 36\fs5$  & $+42\degr 37\arcmin 35\arcsec$ &$4500\pm540$&$2000\pm200$&$ 730\pm60  $\\
DR21~(OH)\tablenotemark{d} & $20\h 39\m 00\fs9$  & $+42\degr 22\arcmin 38\arcsec$&$9100\pm1400$&$4600\pm660$&$2540\pm330 $\\
DR21\tablenotemark{d} & $20\h 39\m 01\fs1$  & $+42\degr 19\arcmin 43\arcsec$&$8800\pm1400$&$3900\pm550$&$ 920\pm120 $\\
Mrk~231         & $12\h 56\m 14\fs23$ & $+56\degr 52\arcmin 25\farcs2$ &$ 6.0\pm0.8$&$ 1.5\pm0.3$& \nodata\tablenotemark{e}     \\
Arp~220\tablenotemark{f} & $15\h 34\m 57\fs21$ & $+23\degr 30\arcmin 09\farcs5$ &$24.2      $&$ 9.8      $&$ 3.9       $
\enddata
\tablecomments{Flux densities and associated uncertainties for BLAST05 targeted
sources in the 250, 350, and 500\,\micron\ bands.  Quoted flux densities
have been color-corrected.  The uncertainties include the estimated contributions
from calibration uncertainty as well as instrumental noise.}
\tablenotetext{a}{Positions are nominal as given by SIMBAD.}
\tablenotetext{b}{Pallas flux densities are average values from 4 observations
taken at the following Julian Dates: 2453534.26; 2453534.29; 2453535.22; and
2453535.40.}
\tablenotetext{c}{The K3-50 region is resolved into two sources by BLAST05.}
\tablenotetext{d}{The W~75N region is resolved into three sources by BLAST05.}
\tablenotetext{e}{Mrk~231 was not detected at 500\,\micron\ by BLAST05.}
\tablenotetext{f}{Arp~220 is the absolute flux calibrator for BLAST05.
Flux densities presented here should thus be considered as
predictions based on a model fit to data from other instruments.
Uncertainties in this model are given in \S~\ref{sec:arp220}}.
\end{deluxetable}

\begin{deluxetable}{lrrrrrrr}
\tabletypesize{\scriptsize}
\tablecaption{Single Temperature SED Best Fit Parameters of BLAST05 Targeted Sources\label{fits}}
\tablewidth{0pt}
\tablehead{
\colhead{Name} & \colhead{$T$} & \colhead{$\beta$} & \colhead{$S_{\rm FIR}$} &
\colhead{Distance} & \colhead{$L_{\rm FIR}$} & \colhead{$M_{\rm dust}$} \\
 & \colhead{[K]} & & \colhead{[W m$^{-2}$]} & \colhead{[kpc]} & \colhead{[L$_\odot$]}
 & \colhead{[M$_\odot$]}
}
\startdata
Pallas\tablenotemark{a}          & $50.2$ & $0.2 \pm 0.2$ & $1.2\E{-12}$ &
                \nodata & \nodata & \nodata \\
CRL~2688\tablenotemark{a}        & $210$ & $0.4 \pm 0.2$ & $1.2\E{-9}$ &
                  1.25    & $1.7\E{5}$  & $3.2\E{-3}$ \\
LDN~1014        & $12 \pm 3$ & $1.8 \pm 0.5$ & $(3.3 \pm 1.2)\E{-13}$ &
                  0.2     & $0.38 \pm 0.14$  & $(3.7 \pm 3.5)\E{-3}$ \\
IRAS~20126+4104\tablenotemark{a} & $41$ & $1.1 \pm 0.2$ & $9.3\E{-11}$ &
                  1.7     & $4.0\E4$ & $0.67$ \\
IRAS~21078+5211 & $33.7 \pm 2.0$ & $1.0 \pm 0.2$ & $(1.2 \pm 0.1)\E{-10}$ &
                  1.65    & $(1.0 \pm 0.1)\E4$ & $0.29 \pm 0.05$  \\
IRAS~21307+5049 & $32.9 \pm 2.5 $ & $1.2 \pm 0.2$ & $(1.0 \pm 0.1)\E{-11}$ &
                  3.6     & $(4.3 \pm 0.5)\E3$  & $0.11 \pm  0.03$ \\
IRAS~22134+5834 & $23.5 \pm 1.0 $ & $2.1 \pm 0.4$ & $(2.0 \pm 0.8)\E{-11}$ &
                  0.90    & $520 \pm200$ & $0.11  \pm    0.07$   \\
IRAS~23011+6126 & $27.8 \pm 2.8 $ & $0.6 \pm 0.2$ & $(4.3 \pm 0.5)\E{-12}$ &
                  0.73    & $71 \pm 9$  & $0.01  \pm 0.007$\\
K3-50A          & $39.7 \pm 2.8$  & $1.7 \pm 0.2$ & $(6.2 \pm 0.5)\E{-10}$ &
                  8.5     & $(1.4 \pm 0.1)\E6$    & $8.1 \pm 1.4$      \\
K3-50C          & $9.3 \pm 1.6$  & $3.9 \pm 0.5$ & $(2.3 \pm 0.8)\E{-11}$ &
                  8.5     & $(5.2 \pm 2.0)\E4$ & $(1.2 \pm 1.1)\E3$ \\
W~75N           & $65$\tablenotemark{b} & $0.8 \pm 0.2$ & $(2.5 \pm 1.3)\E{-9}$ &
                  3       & $(7.1 \pm 3.6)\E{5}$ & $1.0 \pm 0.1$ \\
DR21~(OH)       & $40$\tablenotemark{c}  & $0.5 \pm 0.3$ & $(7.3 \pm 3.3)\E{-10}$ &
                  3       & $(2.1 \pm 0.9)\E5$ & $4.9 \pm 0.7$  \\
DR21            & $29.2\pm3.0$ & $2.3 \pm 0.2$ & $(1.7 \pm 0.3)\E{-9}$ &
                  3       & $(4.9 \pm 0.8)\E5$ & $8.7 \pm 3.0$ \\
Mrk~231         & $43.9 \pm 2.5$  & $1.5 \pm 0.1$ & $(1.8 \pm 0.2)\E{-12}$ &
                  175\,000  & $(1.6 \pm 0.1)\E{12}$ & $(6.4 \pm 0.8)\E{6}$\\
Arp~220\tablenotemark{d}
                & $41.7 \pm 3.5$  & $1.3 \pm 0.1$ & $(5.9 \pm 0.8)\E{-12}$ &
                  75\,000   & $(1.04 \pm 0.14)\E{12}$ & $(7.8 \pm 1.9)\E6$
\enddata
\tablecomments{Parameters for a single temperature modified blackbody fit to
the SED.
$T$ and $\beta$ are the best fit to the BLAST05 and other data as indicated
in the text.  $S_{\rm FIR}$ is the total flux from the modified blackbody fit.
Errors are from 100 Monte Carlo simulations of the fit.  Distances given
are from references indicated in \S~\ref{sec:obs}.  FIR luminosity is
based on these adopted distances.  $M_{\rm dust}$ is based on the formula
$M_{\rm dust} = (S_\nu D^2)/(\kappa B_\nu(T))$,
with an assumed value of 10 for $\kappa$ \citep[see][]{chapin2008}.}
\tablenotetext{a}{The data included in the fit do not accurately constrain the
temperature, so no error bars are given.}
\tablenotetext{b}{Data from \citet{harvey1977} suggest a temperature of 65\,K,
which is used as a constraint in the fit.}
\tablenotetext{c}{Data from \citet{harvey1977} suggest a temperature of 40\,K,
which is used as a constraint in the fit.}
\tablenotetext{d}{Arp~200 is the primary calibrator for BLAST05.  This is our
best-fit to the other data, which we use to calibrate the BLAST photometry.}
\end{deluxetable}

\end{document}